\documentclass[preprint]{aastex631}

\shorttitle{Dyson Sphere Feedback}
\shortauthors{Huston \& Wright}
\graphicspath{{./}{figures/}}

\usepackage{amsmath}
\usepackage{amsfonts}
\usepackage{amssymb}
\usepackage{upgreek}
\usepackage{graphics}
\usepackage{multirow}
\usepackage{}
\defcitealias{mesa1}{\tt MESA}
\defcitealias{mesa2}{{\tt MESA} II}

\begin{document}

\title{Evolutionary and Observational Consequences of Dyson Sphere Feedback}

\author[0000-0003-4591-3201]{Macy Huston}
\affiliation{Department of Astronomy \& Astrophysics, The Pennsylvania State University, University Park, PA 16802}
\affiliation{Penn State Extraterrestrial Intelligence Center, The Pennsylvania State University, University Park, PA 16802}
\affiliation{Center for Exoplanets and Habitable Worlds, The Pennsylvania State University, University Park, PA 16802}

\author[0000-0001-6160-5888]{Jason Wright}
\affiliation{Department of Astronomy \& Astrophysics, The Pennsylvania State University, University Park, PA 16802}
\affiliation{Penn State Extraterrestrial Intelligence Center, The Pennsylvania State University, University Park, PA 16802}
\affiliation{Center for Exoplanets and Habitable Worlds, The Pennsylvania State University, University Park, PA 16802}

\correspondingauthor{Macy Huston}
\email{mzh616@psu.edu}

\begin{abstract}
    The search for signs of extraterrestrial technology, or technosignatures, includes the search for objects which collect starlight for some technological use, such as those composing a Dyson sphere. These searches typically account for a star's light and some blackbody temperature for the surrounding structure. However, such a structure inevitably returns some light back to the surface of its star, either from direct reflection or thermal re-emission. In this work, we explore how this feedback may affect the structure and evolution of stars, and when such feedback may affect observations. We find that in general this returned light can cause stars to expand and cool. Our {\tt MESA} models show that this energy is only transported toward a star's core effectively by convection, so low mass stars are strongly affected, while higher mass stars with radiative exteriors are not. Ultimately, the effect only has significant observational consequences for spheres with very high temperatures (much higher than the often assumed $\sim$300 K) and/or high specular reflectivity. Lastly, we produce color-magnitude diagrams of combined star-Dyson sphere systems for a wide array of possible configurations.
\end{abstract}


\section{Introduction} 
\label{intro}
\cite{dyson60_science} introduced the idea, now referred to as a Dyson sphere or swarm, of large structures surrounding a star and collecting its energy for some intelligent use. A monolithic solid spherical shell could not be built stably \citep{dyson60_letters}, so a collection of many satellites, which \cite{dyson66_essay} demonstrated to be physically feasible, is the preferred model. However, for simplicity of language, we refer to any configuration of a starlight-manipulating megastructure as a Dyson sphere in this work. At the dawn of SETI, when radio searches \citep{cocconi_morrison, ozma} were the primary focus, \citeauthor{dyson60_science} proposed searching for infrared waste heat output from the technological use of starlight. He suggested that these structures may be best suited to orbits similar to that of Earth, where they would be in a temperature range of roughly 200--300 K, making $\sim$10 microns the ideal wavelength range to search.

Past works \citep{badescu95, dsreview} have presented the thermodynamics of Dyson spheres, including radiative feedback from the sphere onto the surface of a star. However, these assume that this irradiation has no significant effect on the interior of stars and thus their structure and evolution. This may not be adequate for accurately predicting the observable features of Dyson spheres that return significant feedback to their stars. 

\subsection{Expectations for Irradiated Stars}
We begin with a look at the general effects of irradiation on stellar structure and evolution. Stars have negative gravitothermal heat capacity: the addition of energy causes them to expand and cool. In a simple theoretical demonstration, the total energy of a star is the sum of its thermal and gravitational energies:
\begin{equation}
    E_{*} = E_{\rm therm} + E_{\rm grav} .
\end{equation}
We apply the virial theorem,
\begin{equation}
    E_{\rm therm} = -\frac{1}{2} E_{\rm grav},
\end{equation}
to relate each energy type directly to the total energy:
\begin{equation}
    E_{*} = \frac{1}{2} E_{\rm grav} = - E_{\rm therm}.
\end{equation}
When energy is added to a star ($E_*$ increases), gravitational energy increases and thermal energy decreases, so we see the star expand and cool both overall (because $E_{\rm therm}$ is lower) and on its surface (because, being larger at the same or a lower luminosity its effective temperature must drop). A larger star should also result in less pressure on a cooler core, so we also expect its luminosity to decrease.

To examine this effect quantitatively, we calculate the differences in radius and temperature between normal and irradiated stellar models. Foundational work on the irradiated stars was performed by \cite{tout89}, who modelled the evolution up to the helium flash of 0.5, 0.8, 1, and 2 M$_\odot$ stars in constant temperature radiation baths from 0 to 10$^4$ K. \citeauthor{tout89} explore their 1M$_\odot$ model in great detail, finding that in the early main-sequence in the 10$^4$ K bath, the star's radius increases significantly (by a factor of 1.4) and that its central temperature and nuclear luminosity decrease very slightly ($<$1\% and $\sim$1\% respectively). They suggest that this large radius increase appears to be caused by the outer convective region becoming radiative. Since a radiative temperature gradient is smaller than a convective gradient, the envelope must be larger to encapsulate a similar temperature drop. The lower luminosity results in a slightly extended main sequence lifetime ($\sim$2\% for the 10$^4$ K bath). 

Since normal 2 M$_\odot$ models have radiative exteriors that are hotter to begin with, \citeauthor{tout89} find that they are not significantly affected by irradiation while on the main sequence. The irradiated 0.8 M$_\odot$ models, normally with an outer convective zone and radiative core, behave similarly to the 1 M$_\odot$ models, expanding and cooling in the outer regions, but not strongly affecting their main sequence lifetimes. Interestingly, \citeauthor{tout89} find that the 10$^4$ K irradiation decreased the main sequence lifetime of their 0.5 M$_\odot$ star by roughly half. They found the star to expand and cool overall, but a radiative region formed, and the thermal energy was distributed in such a way that the central temperature actually increased. 

\subsection{Radiative Feedback from Dyson Spheres}
There are two primary ways that a star may undergo feedback from its surrounding sphere: materials may directly reflect some amount of starlight back onto the star, and/or the sphere may become warm and emit thermally. In the classic idea of a Dyson sphere, the sphere's goal is to collect energy, not to reflect it back. As the sphere collects energy, it will heat up and inevitably emit thermally. Some of this emission may end up going back toward the star, so we can model this process as the diffuse reflection of some fraction of starlight.

Although not spherically symmetrical like the typical modeling of a Dyson sphere, the related concept of ``stellar engines" proposes a partial sphere which uses a significant portion of a star's luminosity to do mechanical work on the star. First proposed by \cite{shkadov}, this could be in the form of a large mirror placed at some distance from the star, disturbing the radiative symmetry of the radiation field and altering its space velocity. \cite{engines00} noted that this reflection will cause the photospheric temperature to increase and gradually change the star's steady state, but they assumed that the nuclear reaction rate is unchanged. In this scenario, a smooth, mirrored surface reflects back as much light as possible, so we can model it as the specular reflection of some fraction of starlight.

\subsection{Dyson Sphere Searches}
As originally suggested by \citeauthor{dyson60_science}, and expanded upon by \cite{sagan_walker}, the search for Dyson spheres began as a search for infrared sources in the Galaxy, assuming that they would appear as blackbodies with a temperature of a few hundred Kelvins. The {\it IRAS} survey revealed the presence of many infrared sources in the galaxy, but the issue remains of how to distinguish between artificial structures and natural objects of similar mid-infrared fluxes, such as young stars with circumstellar disks, or giant stars with dusty atmospheres.  Several searches for Dyson spheres with \textit{IRAS} were performed. \cite{iras1985} and \cite{iras2000} each found several objects with SEDs resembling that of a blackbody with effective temperature between 3-400 K. Each noted that further observations would be needed to rule out natural sources. A series of studies \citep{jugaku95, jugaku97, jugaku00, jugaku2004} combined {\it IRAS} and {\it 2MASS} data to search for partial dyson spheres around solar-type (FGK) stars within 25 pc of the Sun; they found no candidates in their 384 stars studied. 
\cite{iras2009} explored {\it IRAS} Low Resolution Spectrometer (LRS) sources with temperatures below 600 K for Dyson spheres, finding 16 candidates in a study sensitive to solar luminosity objects out to 300 pc. \citeauthor{iras2009} noted that most of these candidates have possible non-technological explanations and that more information is needed to rule out or strengthen the case for any of them.

\cite{teodorani} proposed a more modern approach to the infrared search for Dyson spheres than could be done with {\it IRAS}, using the higher-resolution {\it Spitzer} space telescope to search for sun-like stars with infrared excess. This proposed method, however, assumes that only 1\% of total starlight is obscured, as it relies on optical detectability. \cite{ghat2} laid out the AGENT formalism for describing Dyson spheres and mentioned backwarming as a complication in the prediction of star-Dyson sphere system SEDs. Their {\it WISE} search focused on Kardashev Type III civilizations, or galaxies full of Dyson spheres, and concludes that they are rare or non-existent. \citeauthor{ghat2} also note that infrared waste heat is much easier to detect than blocked optical starlight, except in cases where a very large fraction of the starlight is blocked.

Observing a lack of or change in optical light has also been the focus of many Dyson sphere searches. One such method is looking for strange light curves that could be indicative of transiting megastructures (\cite{arnold, teodorani, ghat3}). \cite{zackrisson18} proposed and demonstrated the use of {\it Gaia} to search for optically underluminous stars, by identifying those with spectrophotometric distance estimates far larger than parallax distances. 
When {\it JWST} is available for use, \cite{whitepaper} propose that it, in combination with {\it WISE} and {\it Gaia}, can be used to put robust upper limits on stellar energy collection throughout our Galaxy and across others.

\subsection{This Work}

In this work, we use {\tt MESA} \citep{mesa1, mesa2, mesa3, mesa4, mesa5} to model the effects of Dyson sphere feedback on the structure and evolution of stars. We use the AGENT formalism of \cite{ghat2} and radiative feedback formulation of \cite{dsreview} to compute absolute magnitudes of many configurations of combined star-Dyson sphere systems. We produce color-magnitude diagrams for selected {\it Gaia} and {\it WISE} bands and specify when these internal changes in the stars are significant in observations.

Section \ref{sec:mesa} describes the {\tt MESA} modelling methods used in this work. Section \ref{sec:irrad} describes our recreation of the irradiated stellar models of \cite{tout89} in {\tt MESA}. Sections \ref{sec:struc} and \ref{sec:evol} discuss the impacts of the feedback model on stellar structure and evolution. In section \ref{sec:obs}, we demonstrate the observational consequences of these effects. Section \ref{sec:concl} concludes the work and motivates future Dyson sphere searches.

\section{{\tt MESA} Techniques} \label{sec:mesa}

{\tt MESA} \citep[Modules for Experiments in Stellar Astrophysics;][]{mesa1, mesa2, mesa3, mesa4, mesa5} is an open source tool for stellar physics calculations. We use the {\tt MESAstar} module to simulate stellar evolution under the presence of irradiation. The program's adaptive timesteps allow for the thermal processes of external irradiation to occur in appropriate intervals. {\tt MESA} provides three methods for irradiation \citep{mesa2}. Method 1 applies irradiation in the form of energy injection at some specified depth. Method 2 alters the model's boundary conditions in a manner appropriate for irradiated giant planets. Method 3 is similar to 1 but allows for more flexibility with the use of custom energy injection routines, adding any amount of energy at any cell in the 1-dimensional model. The \citetalias{mesa2} paper demonstrates that these three methods produce equivalent results in the case of irradiated planets. Because it is the most flexible of the three methods (and because Method 2 is only appropriate for planets) we adopt the third method, using the {\tt other\_energy} module to deposit irradiation into the outermost cell of the model star.

\subsection{Stars in Constant Temperature Baths}
We first attempt to recreate the results of \cite{tout89} in MESA. We run simulations of 0.5, 0.8, 1, and 2 M$_\odot$ stars in temperature baths from 0 to 10$^4$ K. This temperature bath is implemented using the {\tt run\_star\_extras} function {\tt other\_energy}, which continually pours energy onto a star. For a constant temperature bath, this is implemented as,
\begin{equation}
L_{\rm extra} = 4 \pi \sigma R_*^2 T_{\rm b}^4 ,
\end{equation}
where $L_{\rm extra}/dm$ is the input to {\tt MESA}'s {\tt other\_energy} function (in units of luminosity/mass), $\sigma$ is the Stefan-Boltzmann constant, $R_*$ is the star's radius, $T_{\rm b}$ is the bath temperature in Kelvin, and $dm$ is the mass contained in the outermost cell in our model.

\subsection{Pre-Dyson Sphere Models}
Our {\tt MESA} models begin with the creation of zero age main sequence (ZAMS) models for a range of stellar masses. For stars of mass $\leq 1M_\odot$, we allow evolution along the main sequence for 4.6 Gyr before applying the feedback of a Dyson sphere. For stars with mass $> 1M_\odot$, we evolve to half of the star's main sequence lifetime (in the absence of added feedback).

\subsection{Calculating Dyson Sphere Feedback}
To replicate the radiation of a Dyson Sphere, we pour some fraction of the star's luminosity back onto the surface of the star.
Physically, the extra luminosity is a simple function of the star's luminosity,
\begin{equation}
L_{\rm extra} = f L_*,
\end{equation}
where $L_{\rm extra}/dm$ is the input to {\tt MESA}'s {\tt other\_energy} function (in units of luminosity/mass), $L_*$ is the star's luminosity, $f$ is the fraction of the star's luminosity reflected back on itself, and $dm$ is the mass contained in the outermost cell in our model.

However, {\tt MESA}'s finite and relatively long timesteps add some complications, so we have to add some corrections into our function. Without any feedback, the luminosity of the outer layer ($L[1]$) will be equivalent to that just below it ($L[2]$), so we start with,
\begin{equation}
    L_{\rm extra} = f L[1] = f L[2] .
\end{equation}
When Dyson sphere feedback hits the star, its surface heats up, and in steady state the surface luminosity increases to:
\begin{equation}
    L[1] = (1+f) L[2],
\end{equation}
which then increases feedback:
\begin{equation}
    L_{\rm extra} = f (1+f) L[2] .
\end{equation}
This again increases surface luminosity,
\begin{equation}
    L[1] = (1+f(1+f)) L[2],
\end{equation}
which increases feedback:
\begin{equation}
    L_{\rm extra} = f(1+f(1+f)) L[2].
\end{equation}
This feedback cycle happens much quicker than the {\tt MESA} timescale, so we implement a correction factor which extends this cycle to infinity, following the limits taken in Equations  26 and 27 of \cite{dsreview}. This results in a feedback luminosity of
\begin{equation}
    L_{\rm extra} = \left( \sum_{i=0}^{\infty} f^i \right) L[2] = \frac{1}{1-f} L[2] .
\end{equation}
In essence, because the ``reflection'' of light between the star and the sphere happens much faster than an evolutionary timestep in MESA, we have just treated the sphere and star as partial mirrors within a timestep. In the case of $f=1$, the energy is completely trapped between the sphere and star and so $L_{\rm extra}$ diverges (because the number of reflections does).

We also need to correct for stellar luminosity evolution within a timestep; we found that without such a correction our simulations under-irradiated the star by potentially significant amounts. We forecast our next step luminosity as the current step's plus the difference between the current step and the one before it. So, we replace $L[2]$ with $2L[2] - L_{\rm prev}[2]$. Thus, our final input for {\tt other\_energy} is:
\begin{equation}
    L_{\rm extra} = \frac{f}{1-f}(2L[2] - L_{\rm prev}[2]).
\end{equation}
We found that for large values of $f$, this feedback could become unstable and generate large oscillations in the radiative feedback. To correct this, we also implement a gradual onset of the feedback over multiple steps for $f>0.25$ cases to prevent instability. We checked the fidelity of our implementation of this model in the MESA simulations (all with $f \leq 0.50$) by confirming that $|L_{extra}-fL[1]|<10^{-5}$.

\subsection{Final Model Parameters}
We evolve stars with masses 0.2, 0.4, 1, and 2 $M_\odot$ under the effects of Dyson sphere feedback, with a set of different feedback levels from 1-50\%. We pick up where the mid-main sequence models described above left off and evolve each star to the end of its main sequence phase. All of these models include the repeated feedback correction. Due to a numerical stability issue, all models except for the 50\% feedback case for 0.2 and 0.4 M$_\odot$ stars include the luminosity evolution correction.

\section{Consistency with Prior Work} \label{sec:irrad}
\begin{figure}
    \centering
    \includegraphics[width=0.47\textwidth]{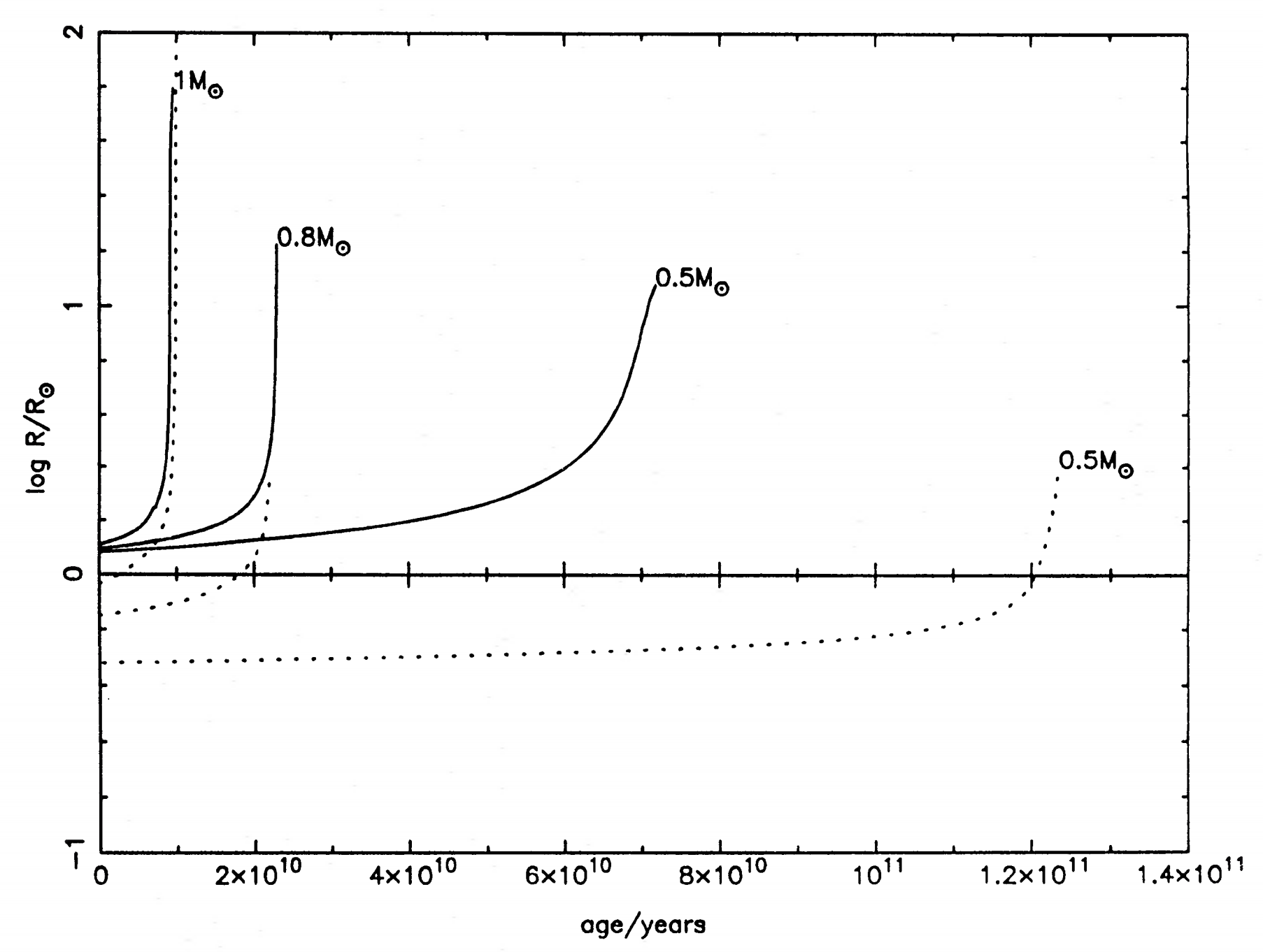}
    \includegraphics[width=0.51\textwidth]{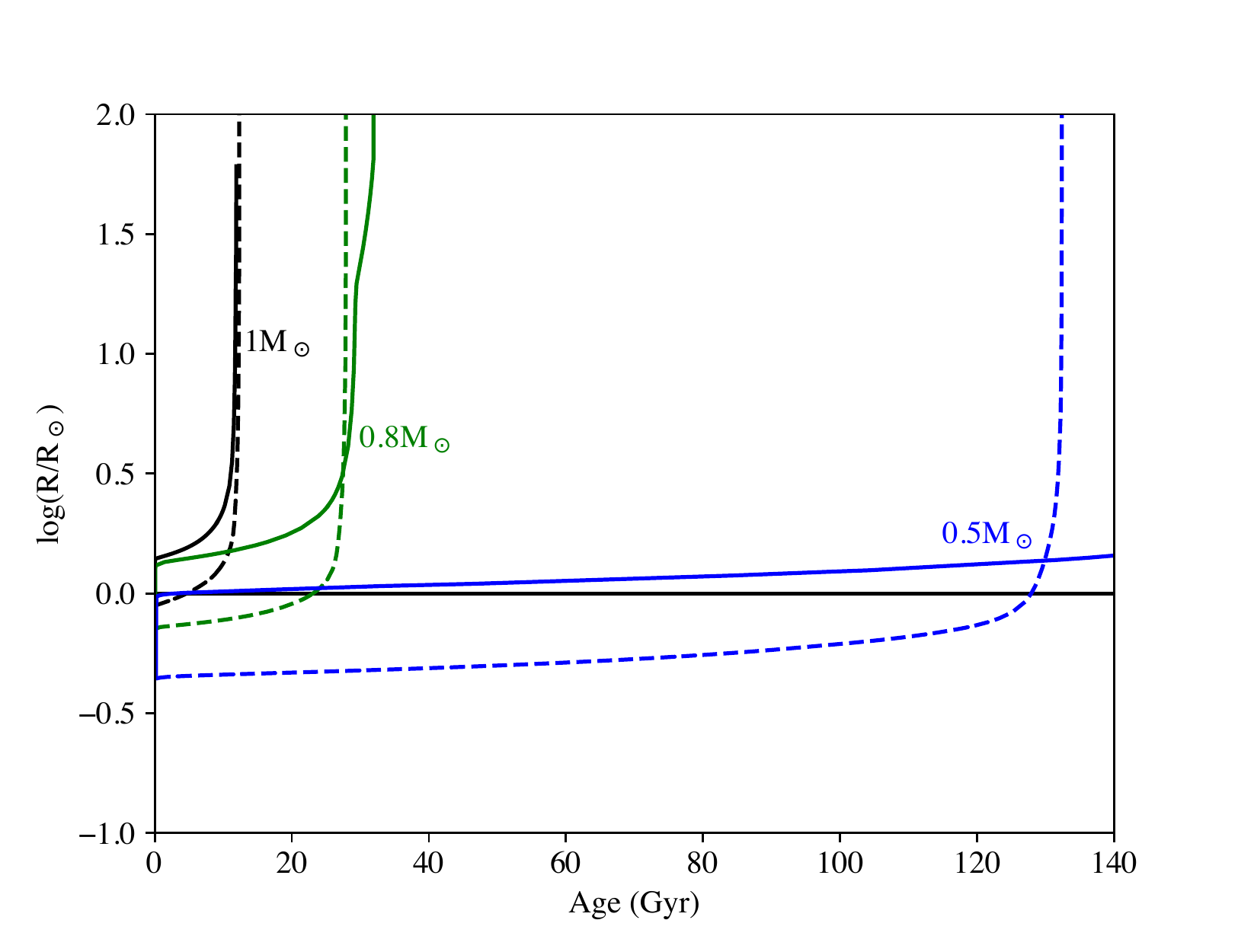}
    \caption{The evolution of stellar radius over time for 0.5, 0.8, and 1 M$_\odot$ stars over time, for a normal star and one in a 10,000 K bath. Left: Figure 5 from \cite{tout89} (solid lines: 10,000 K bath; dotted lines: normal star). Right: Our recreation with MESA. The 0.8 and 1 M$_\odot$ stars evolve very similarly between the two figures. The irradiated 0.5M$_\odot$ model, however, survives much longer in our version that that of \citeauthor{tout89}, nearly doubling its normal main sequence lifetime instead of halving it.}
    \label{fig:tout}
\end{figure}

In Figure~\ref{fig:tout}, we compare our constant-temperature bath irradiated stars with those of \cite{tout89} to check for consistency. Our 1M$_\odot$ \citetalias{mesa1} models see a similar shrinking of the main sequence convective zones and that they almost entirely vanish for the 10$^{3.94}$ and 10$^4$ K bath models. This results in a slightly extended main sequence lifetime ($\sim$2\% for the 10$^4$ K bath). We find similar agreement for the 0.8 and 2 M$_\odot$ stars.

Our results, however, disagree significantly for the 0.5 M$_\odot$ star. The {\tt MESA} modelling found the irradiated 0.5 $M_\odot$ star to cool, even in the core, and to increase in lifetime by roughly 60\%. Without full details of the models used by \citeauthor{tout89}, it is hard to say for sure why our results varied. A comparison between our stars' radius evolution and that of \cite{tout89} is shown in Figure \ref{fig:tout}. We suggest that the difference could be due to {\tt MESA}'s having more modern and accurate opacity tables for cool stars than those used in older stellar models.

\section{A Dyson Sphere's Feedback's Impact on Stellar Structure} \label{sec:struc}
We evolve each star from the onset of its Dyson sphere feedback to the end of the main sequence. Along the way, we save a model of each star after it has 100 Myr to settle from the onset of the sphere. In this section, we compare the structures of each model at this stage, focusing on primarily on temperature. In order to view both the core and envelope in detail, we use the {\tt logit$_{10}(m/M_*)$} function to plot mass, where
\begin{equation}
    {\rm logit_{10}}(x) = \log_{10}\left(\frac{x}{1-x}\right).
\end{equation}
\noindent This scaling extends both the center and edge of the star logarithmically, so -10 refers to the innermost $10^{-10}$ of the star by mass, +10 refers to the outermost $10^{-10}$ by mass, and 0 refers to the midway point where $m/M_* = 0.5$.

\subsection{Predominantly Convective Stars}
\begin{figure}
    \centering
    \includegraphics[width=0.9\textwidth]{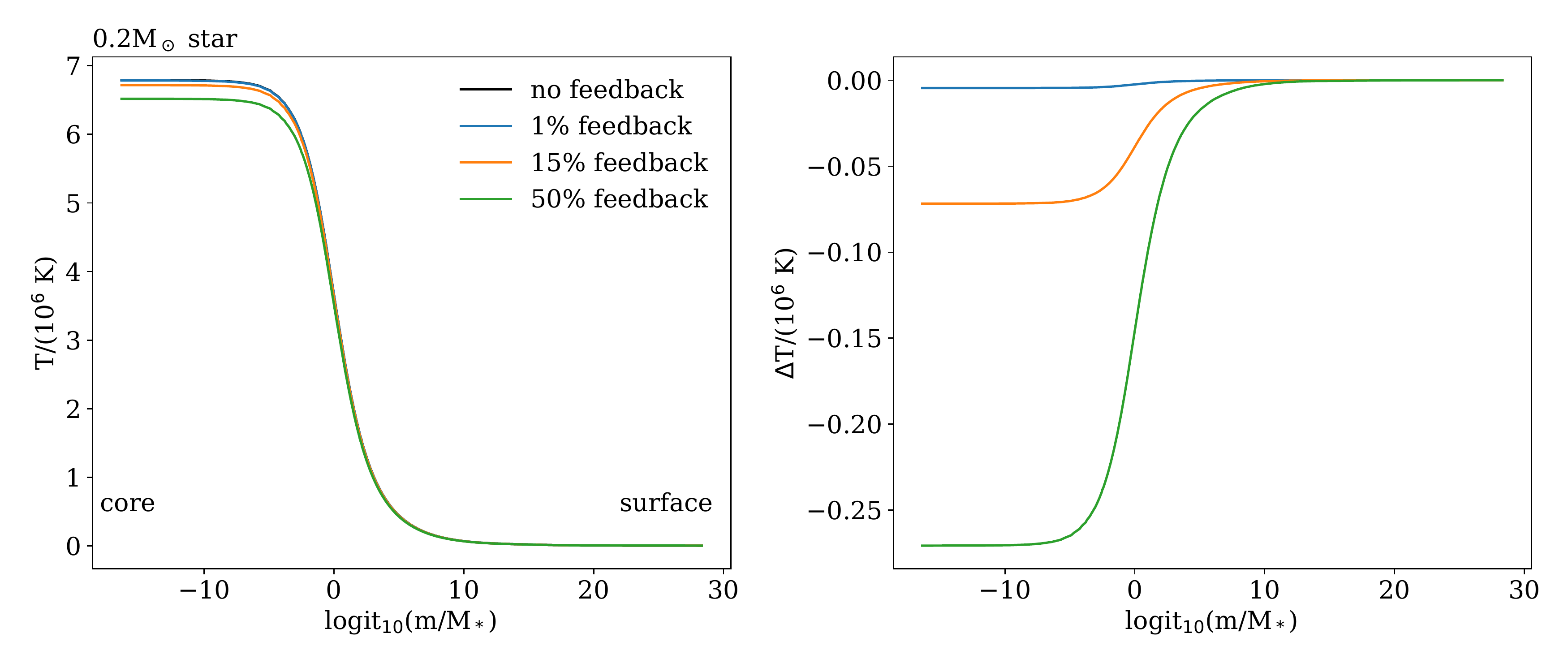}
    \caption{Temperature structure of a 0.2M$_\odot$ star, given 100 Myr to settle after the onset of Dyson sphere feedback. See Section~\ref{sec:struc} for an explanation of the x-axis. Left: Temperature throughout each stellar model. Right: Temperature deviation from the zero feedback model throughout each of the three feedback models. The stars undergoing feedback have lower central temperatures than the ordinary star and nearly identical outer temperatures.}
    \label{fig:struc02}
\end{figure}

Temperature structure of the fully convective 0.2M$_\odot$ star models given 100 Myr to settle after DS feedback onset are shown in Figure \ref{fig:struc02}. We find significant decreases in central temperature in the DS feedback cases, on the order of 10$^5$ K. This internal temperature change dramatically decreases nuclear fusion, with luminosity decreasing by nearly 50\% in the 50\% feedback case. We also see an increase in radius throughout, particularly prominently in the outer half of the star (not shown).

\begin{figure}
    \centering
    \includegraphics[width=0.9\textwidth]{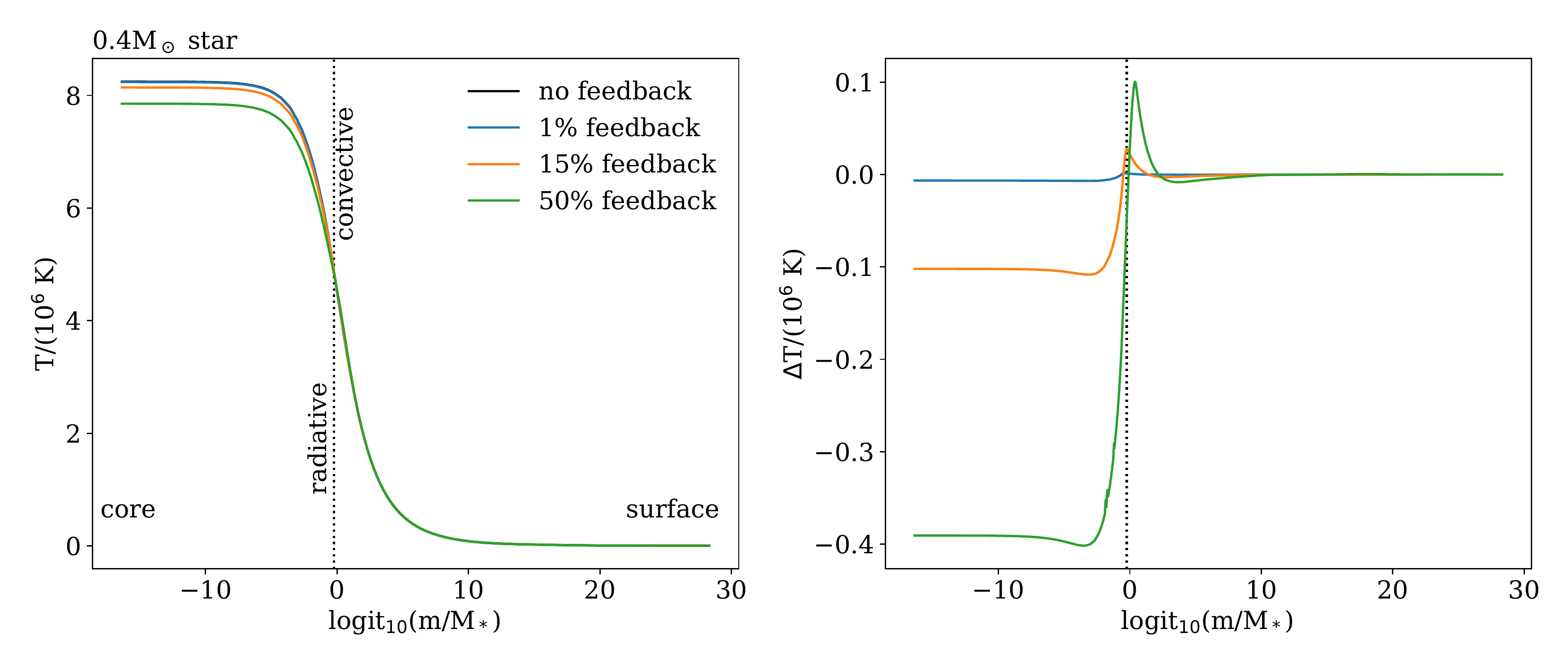}
    \caption{Temperature structure of a 0.4M$_\odot$ star, given 100 Myr to settle after the onset of Dyson sphere feedback. Left: Temperature throughout each stellar model. Right: Temperature deviation from the zero feedback model throughout each of the three feedback models. The dotted line marks the bottom of the convective zone, 0.14M$_\odot$ from the star's center. The stars with feedback again are cooler in their centers and show a temperature at their outer edges. Just above the star's convective boundary, though, we see a spike of higher temperature for these stars.}
    \label{fig:sctruc04}
\end{figure}

Figure \ref{fig:sctruc04} shows the temperature structure of the 0.4M$_\odot$ models 100 Myr after the Dyson sphere onset. This star has a radiative interior containing 35\% of its mass, covered by a convective exterior. The border is shown with a dotted line and is where temperature structure varies the most. We see a spike in temperature for the feedback models just above the bottom of the convective region, where the incoming energy piles up due to inefficient transport into the radiative region below. This is followed by a dramatic dip at the top of the radiative zone, where the irradiation that has made it through has caused expansion and cooling. The spike then settles out toward a central temperature which is reduced by up to hundreds of thousands Kelvin for the stars undergoing feedback. Luminosity decreases for the models with feedback. This effect is less dramatic here than in the 0.2 M$_\odot$ stars, but but it is still significant, decreasing nuclear luminosity by nearly 40\% for the 50\% feedback case.

\subsection{Predominantly Radiative Stars}

\begin{figure}
    \centering
    \includegraphics[width=0.9\textwidth]{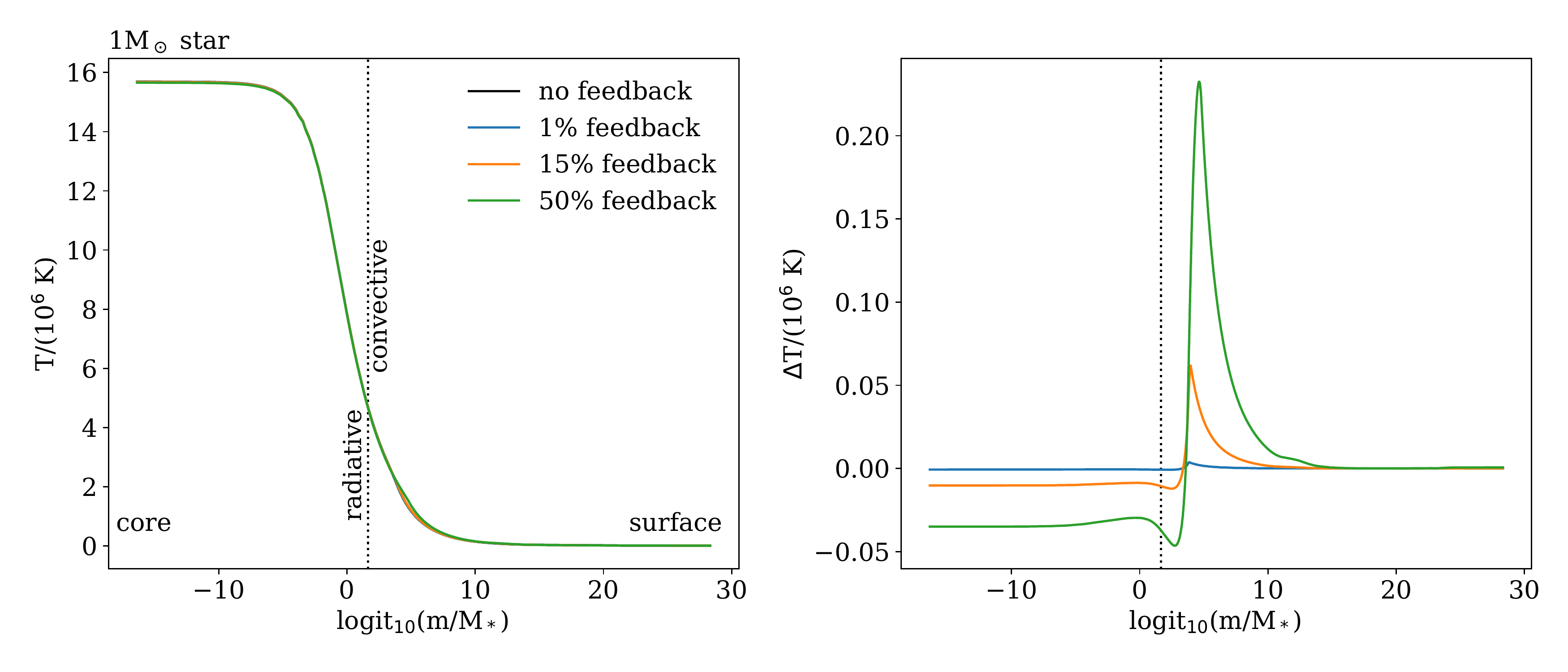}
    \caption{Temperature structure of a 1M$_\odot$ star, given 100 Myr to settle after the onset of Dyson sphere feedback. Left: Temperature throughout each stellar model. Right: Temperature deviation from the zero feedback model throughout each of the three feedback models. The dotted line marks the bottom of the convective zone, 0.98M$_\odot$ from the star's center. The models undergoing feedback are still slightly cooler in the center here but remain the same temperature at their surfaces. Above the convective boundary, we again see a spike in high temperature.}
    \label{fig:struc1}
\end{figure}

The 1 M$_\odot$ star, shown in Figure \ref{fig:struc1}, is primarily radiative, with a light convective exterior containing 2\% of its mass. As the left side of the figure shows, the changes seen are quite small on the scale of the star's actual temperature, but the right panel shows that the small effects are still quite distinctive. Overall we see a similar shape in the temperature differential as for the 0.4M$_\odot$ star, which also has a convective exterior and radiative interior, though with quite different proportions. We see a large spike in increased temperature as one moves toward the convection-radiation border for the models with feedback. Near the bottom of this convective zone, the temperature difference rapidly declines and becomes negative, settling down to a slightly lower central temperature than the normal star. The central temperatures of the 1 M$\odot$ models are much less strongly affected than the 0.4 M$_\odot$ set.
This slight decrease in central temperature produces a very slight decrease in luminosity. We see little expanding and cooling in the radiative interior, but the outer convective zone grows significantly in radius (not shown).

\begin{figure}
    \centering
    \includegraphics[width=0.9\textwidth]{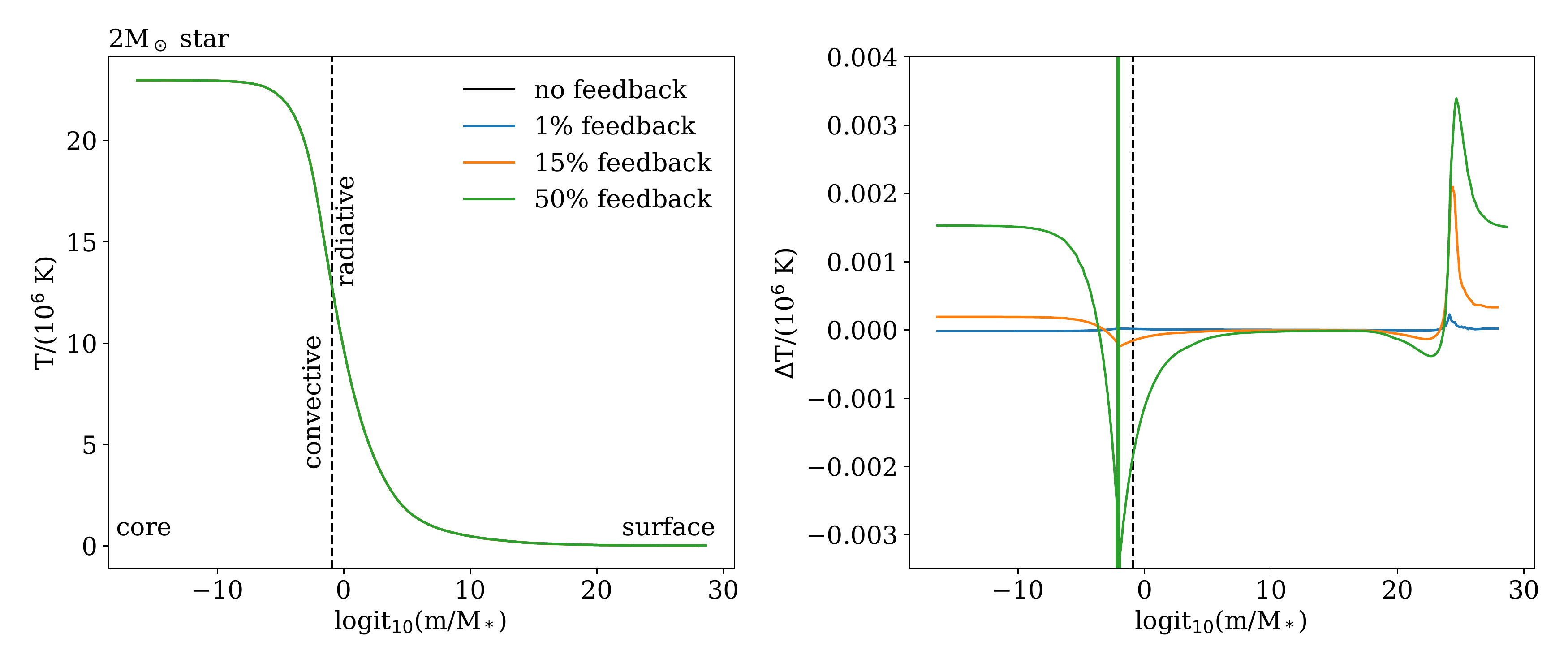}
    \caption{Temperature structure of a 2M$_\odot$ star, given 100 Myr to settle after the onset of Dyson sphere feedback. Left: Temperature throughout each stellar model. Right: Temperature deviation from the zero feedback model throughout each of the three feedback models. The dashed line marks the top of the star's convective interior at 0.21M$_\odot$ from its center. The 50\% feedback model has a numerical artifact just below the convective boundary. We see a very slightly higher internal temperature for the stars with feedback, then a dip to lower temperature at the convective boundary. Moving outward they return to the normal star's temperature, then just at the very outermost edge, they fall below then spike back above the normal temperature.}
    \label{fig:struc2}
\end{figure}

The 2 M$_\odot$ star, shown in Figure \ref{fig:struc2}, has a convective 0.21 M$_\odot$ core, surrounded by a large radiative region. While overall, the left side of the figure shows that the star is not very strongly affected, the right side shows that some points throughout the star have patterns of temperature deviation. At the very surface of the star, we can see that irradiation has increased the surface temperature, though this quickly decreases and goes away as one moves further into the star. This effect is also reflected in the radius (not shown) which is inflated in the very outermost regions of the irradiated stars, then returns to roughly the level of the model with no feedback. Very little of the feedback reaches further into the star, but that which does produces interesting effects surrounding the radiation-convection border.
As one moves deeper into the star approaching this border, temperature begins to decrease, where the nearby convective zone is able to effectively transport heat away. Near the top of this convective zone, the temperature spikes back up, evening out toward a central temperature that is slightly higher for the models with feedback. Also in this region, we see nearly vertical dip and spike in the model with 50\% feedback, which was found to be a numerical artifact. At this turnover point just below the radiative zone, the stars undergoing feedback also see a disturbance in radius (not shown), with a slight decrease before settling out to match up with the feedback-free model. The decreased central temperature is not strong enough to significantly impact nuclear burning. The star's luminosity is essentially unaffected in the core, though the outer region shows very slightly decreased, where fusion is very slightly perturbed at the turnover point below the radiation-convection border.

\section{Feedback's Impact Through Main Sequence Evolution} \label{sec:evol}
\subsection{Predominantly Convective Stars}
\begin{figure}
    \centering
    \includegraphics[width=0.49\textwidth]{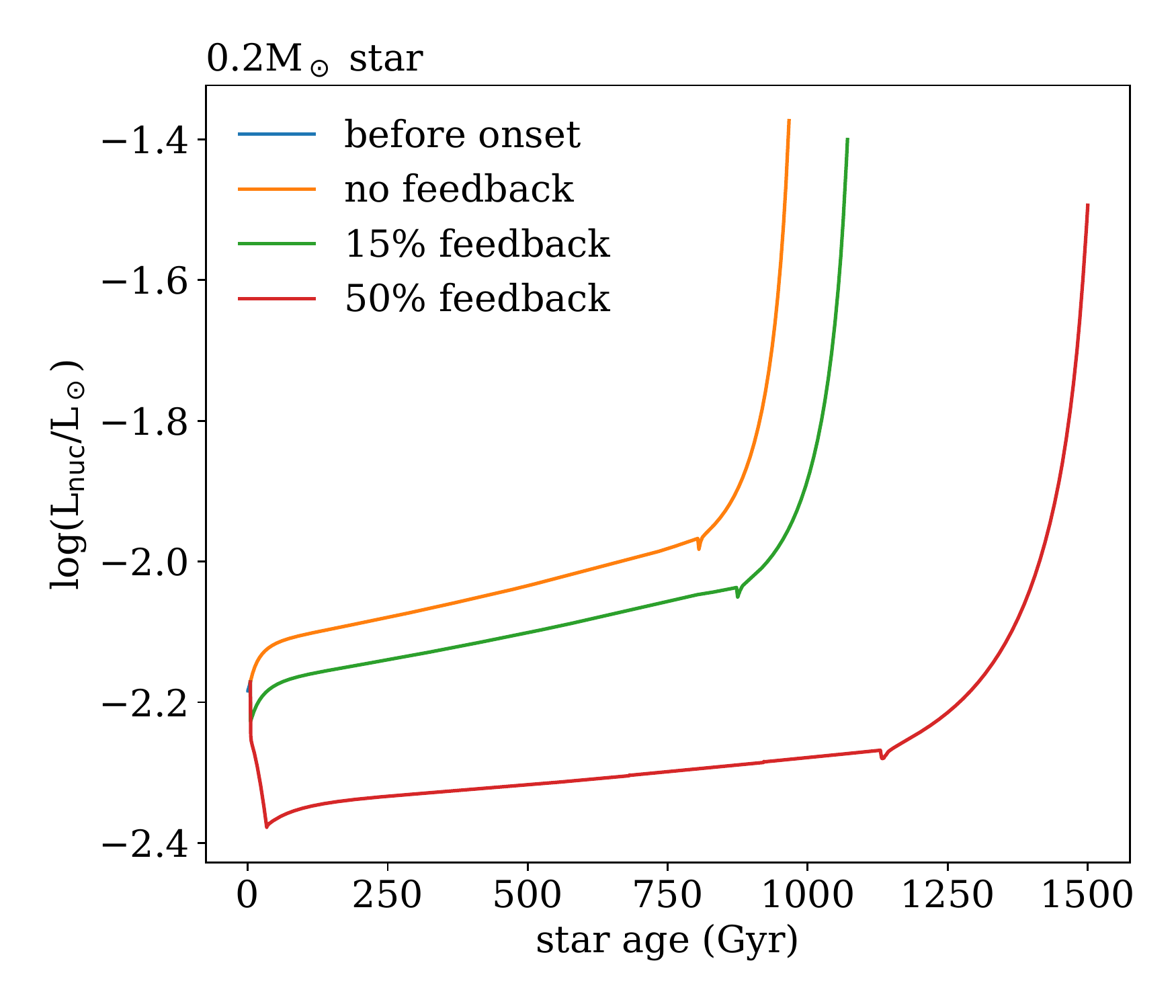}
    \includegraphics[width=0.49\textwidth]{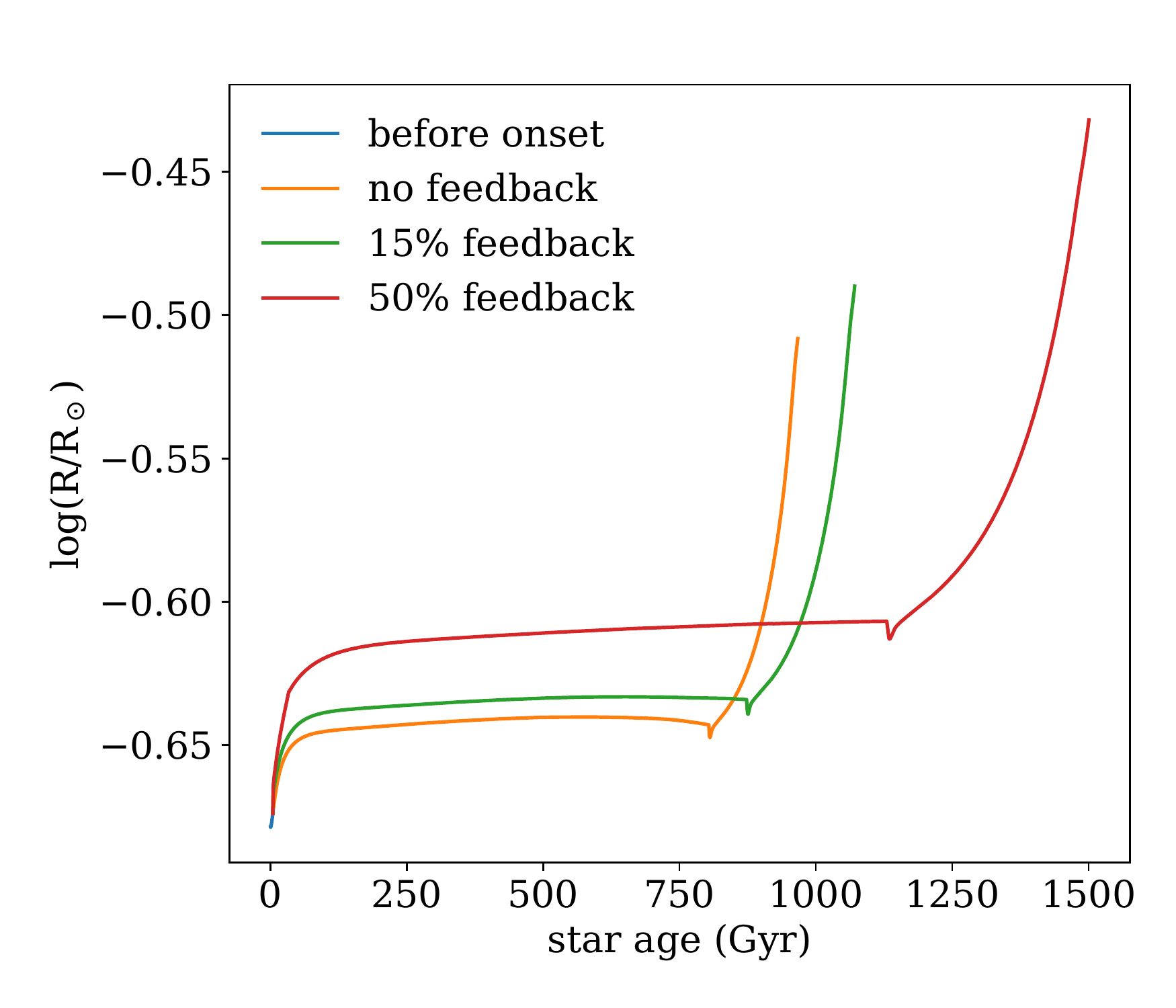}
    \caption{Luminosity (left) and radius (right) evolution of 0.2M$_\odot$ stars for different feedback levels. For stars with significant feedback, nuclear burning dramatically drops, and radius expands. These stars survive longer on the main sequence.}
    \label{fig:evol02}
\end{figure}

Figure \ref{fig:evol02} shows the nuclear luminosity and radius evolution through the main sequence of the 0.2M$_\odot$ models for a few selected feedback levels. The star with 15\% luminosity feedback significantly decreases in nuclear burning and has its main sequence lifetime extended from 970 Gyr to 1070 Gyr. The 50\% feedback case decreases nuclear burning very dramatically and extends the main sequence lifetime to 1500 Gyr, an increase of over $1/3$. We see a significant increase in radius for 15\% feedback and even more so for 50\%. 

\begin{figure}
    \centering
    \includegraphics[width=0.49\textwidth]{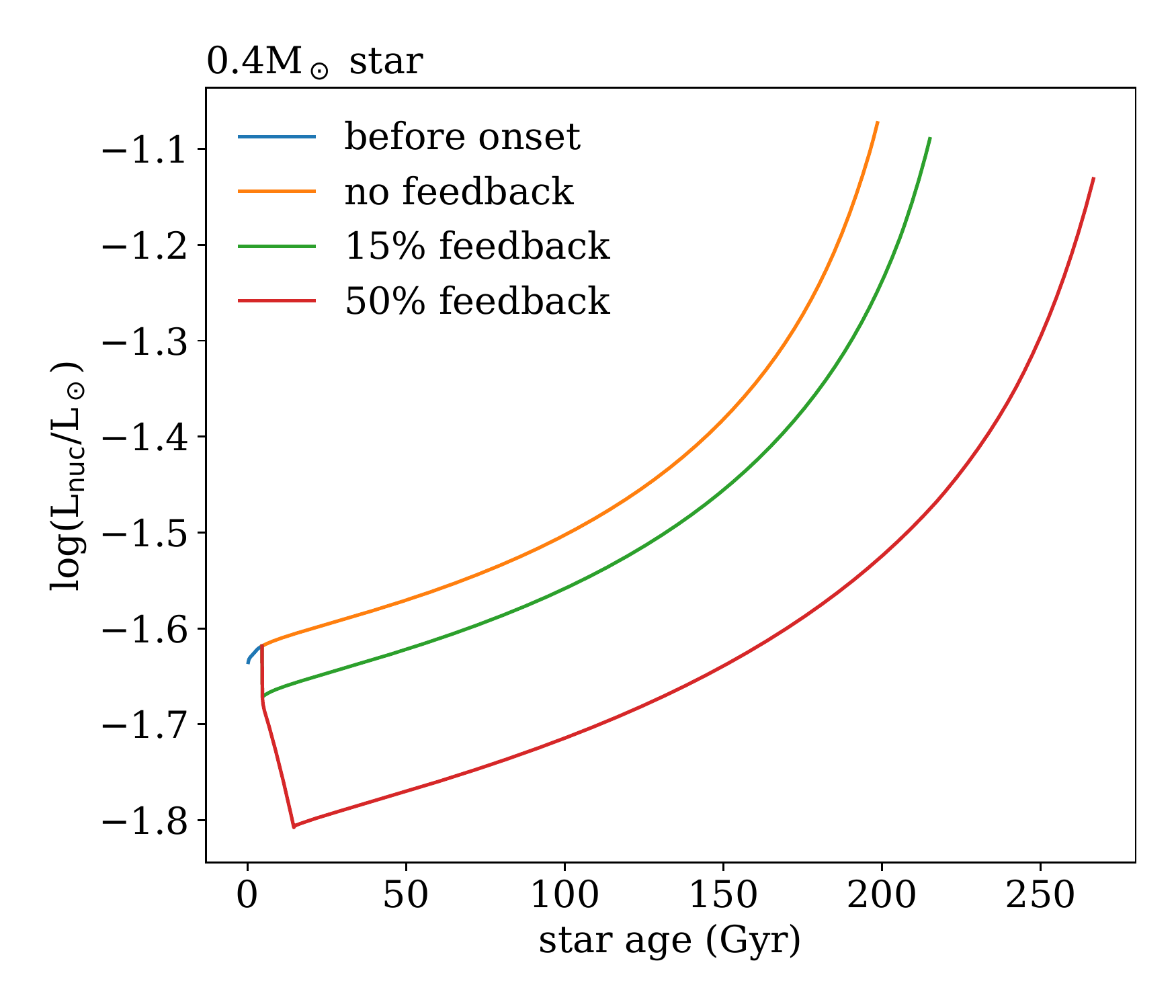}
    \includegraphics[width=0.49\textwidth]{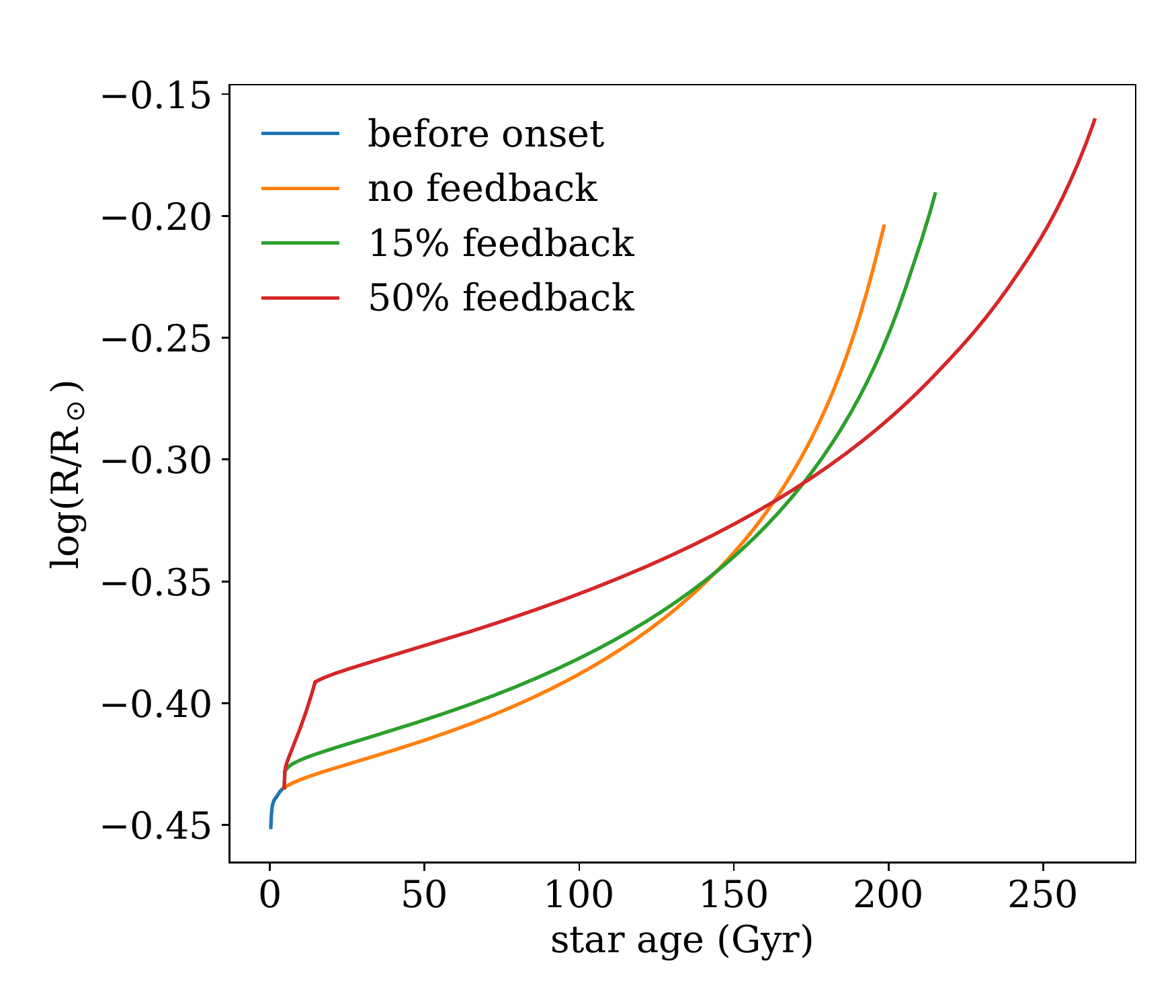}
    \caption{Luminosity (left) and radius (right) evolution of 0.4M$_\odot$ stars for different feedback levels. Nuclear luminosity decreases and radius increases for stars with feedback again, though slightly less dramatically than for the 0.2M$_\odot$ star. Their lifetimes are also extended.}
    \label{fig:evol04}
\end{figure}

We see a similar but slightly less dramatic effect on the 0.4M$_\odot$ models, shown in Figure \ref{fig:evol04}. For higher feedback, nuclear burning decreases and radius increases as the star expands and cools. The 15\% case extends the main sequence lifetime from 199 Gyr to 215 Gyr. The 50\% case has a main sequence lifetime of 267 Gyr.

At the onset of the Dyson sphere feedback, the 0.2M$_\odot$ star is completely convective, while the 0.4M$_\odot$ star is radiative out to 35\% of its mass and convective in the exterior. As predicted, we see a very strong cooling and expanding effect on these highly convective stars, significantly extending their lifetimes. The effect is stronger on a fully convective star than one with a radiative core.

\subsection{Predominantly Radiative Stars}
\begin{figure}
    \centering
    \includegraphics[width=0.49\textwidth]{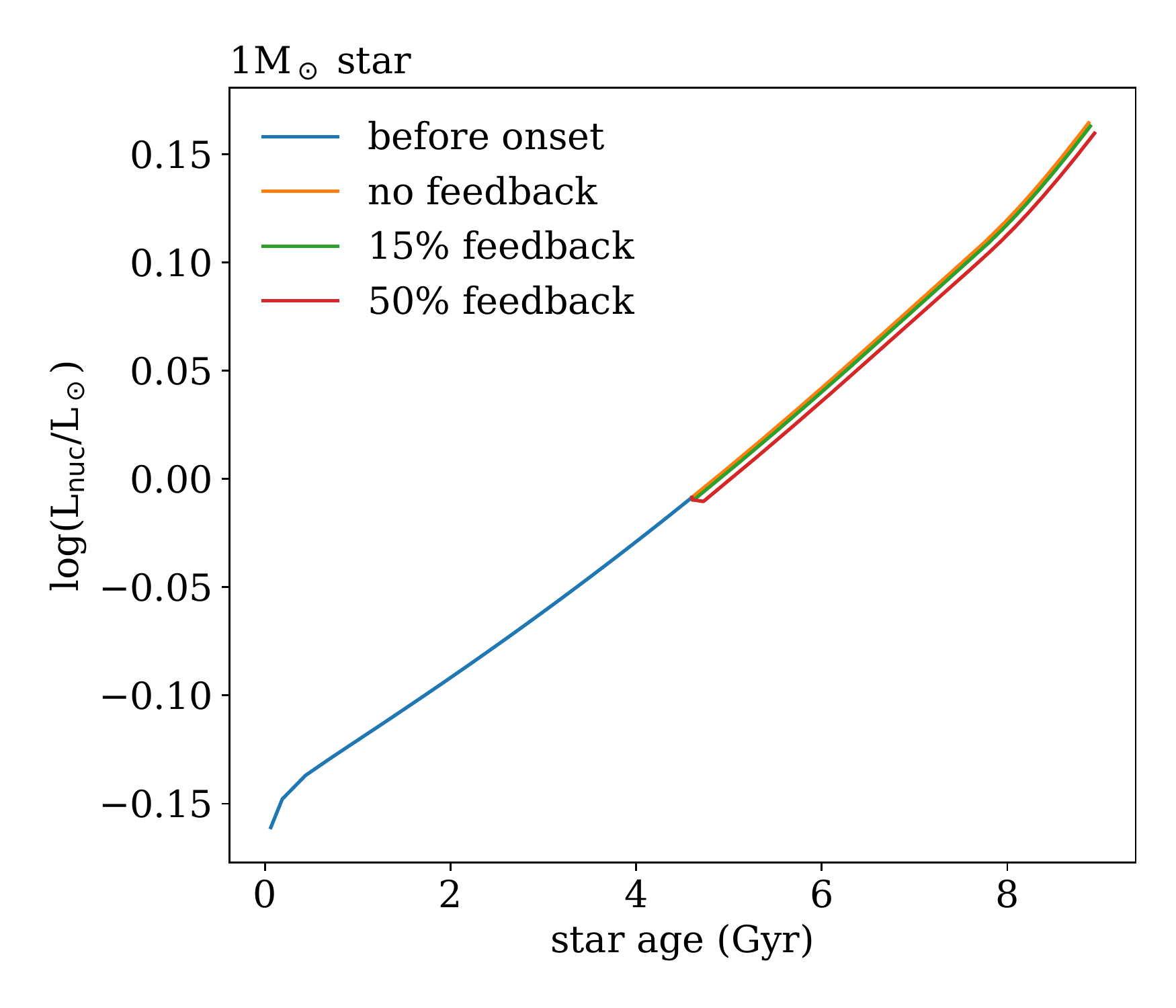}
    \includegraphics[width=0.49\textwidth]{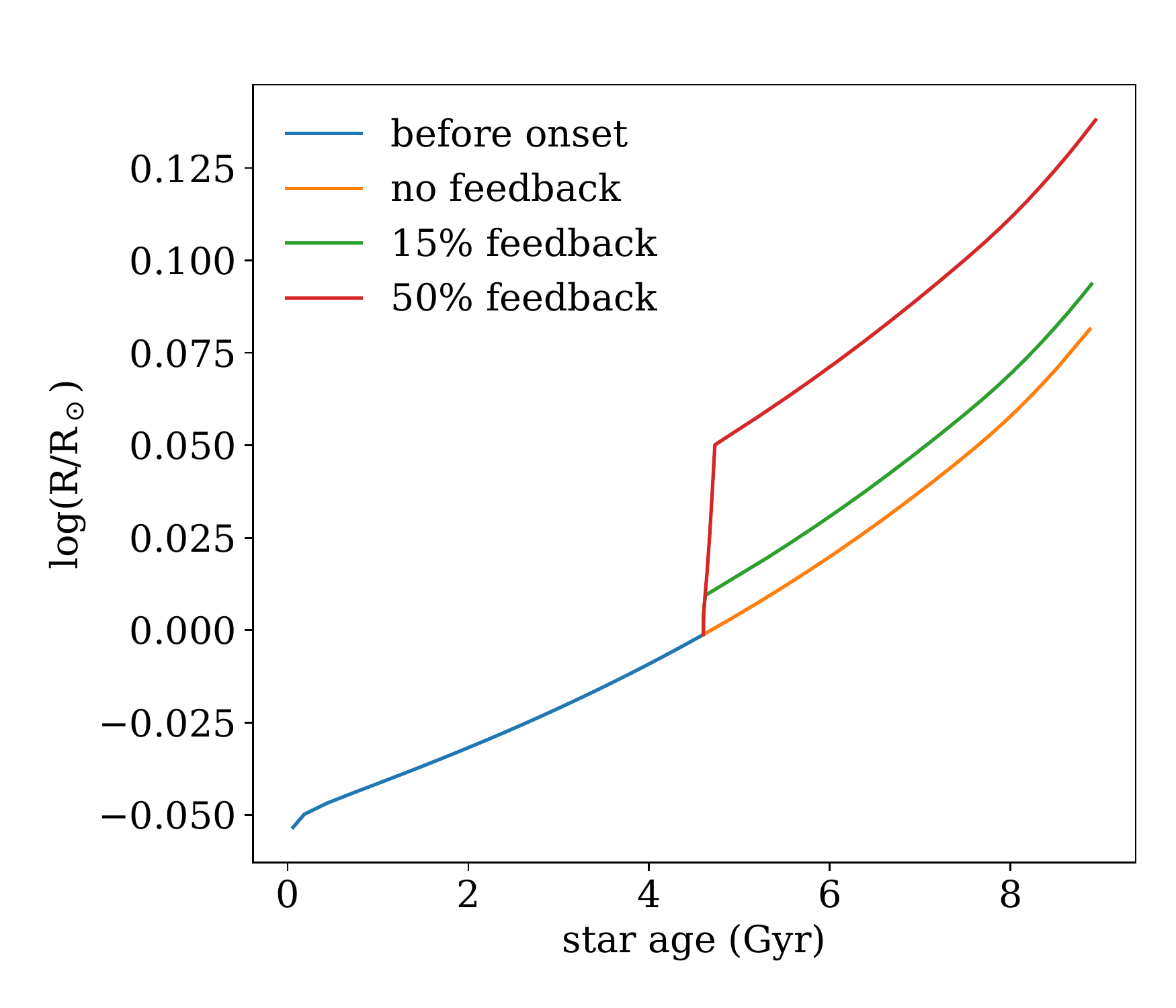}
    \caption{Luminosity (left) and radius (right) evolution of 1M$_\odot$ stars for different feedback levels. Radiative feedback has a large effect on the convective envelope, causing these stars' radii to grow significantly. But, because this envelope has so little of the star's mass, nuclear burning only very slightly decreases, and lifetimes are not significantly effected.}
    \label{fig:evol1}
\end{figure}

\begin{figure}
    \centering
    \includegraphics[width=0.49\textwidth]{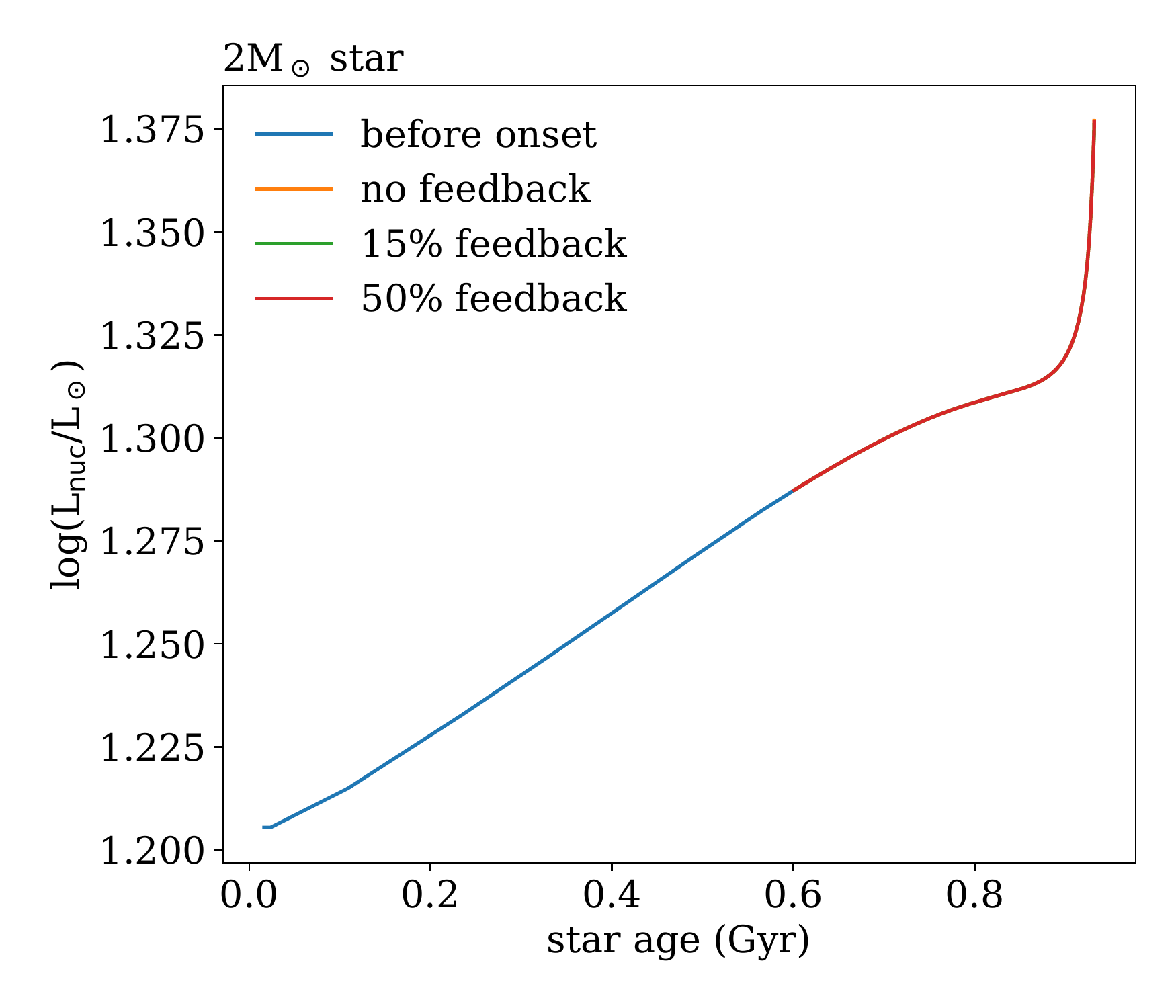}
    \includegraphics[width=0.49\textwidth]{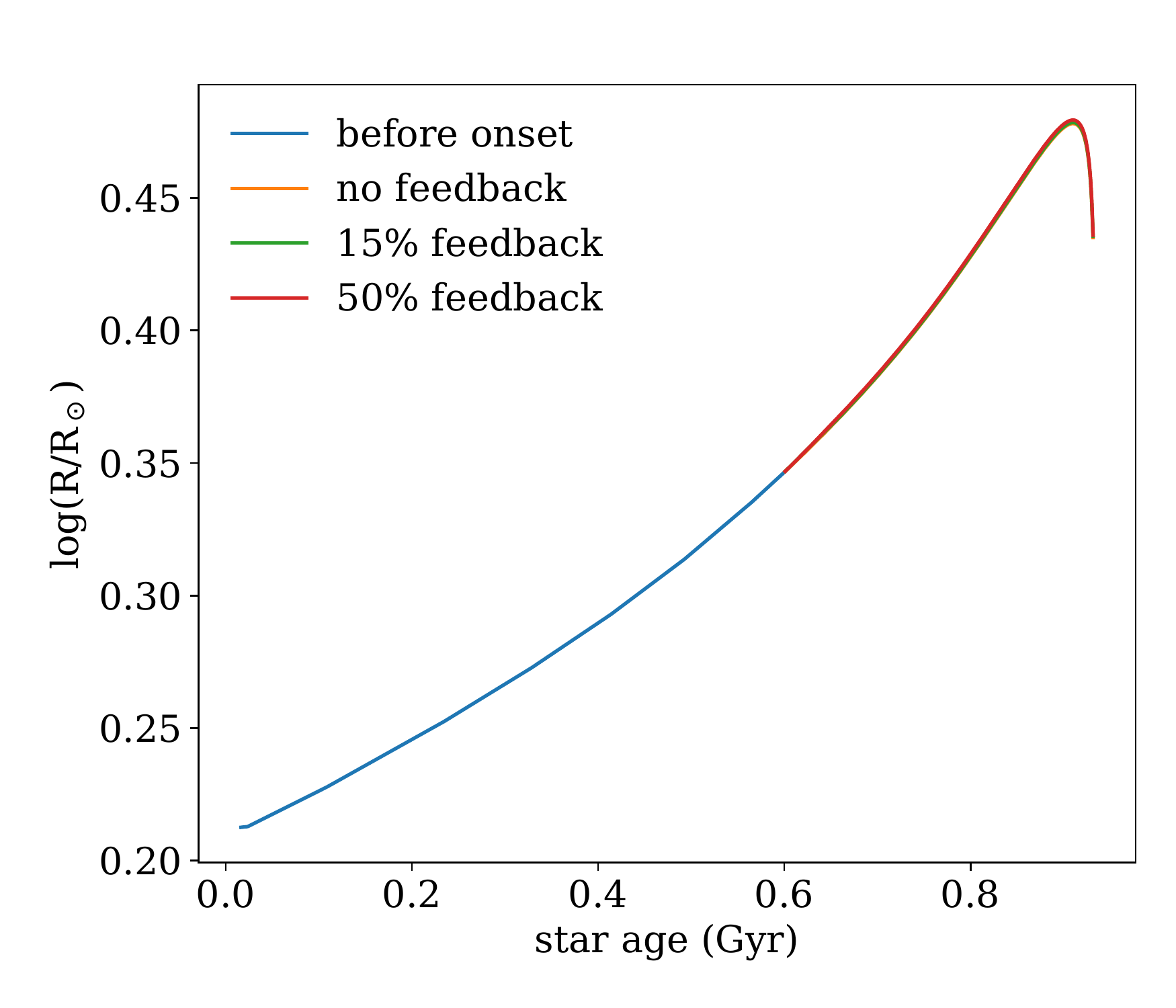}
    \caption{Luminosity (left) and radius (right) evolution of 2M$_\odot$ stars for different feedback fractions. Because this type of star has a radiative envelope, feedback has virtually no effect on the star's bulk properties.}
    \label{fig:evol2}
\end{figure}

Figure \ref{fig:evol1} shows the nuclear luminosity and radius evolution through the main sequence for selected feedback levels on a 1M$_\odot$ star. We see very slight dips in luminosity at the onset of the Dyson sphere. The nuclear burning decreases very slightly, causing lifetimes to be slightly extended, from 8.88 Gyr to 8.90 Gyr for 15\% feedback and 8.94 Gyr for 50\% feedback. Interestingly, in contrast with the minor effects on fusion and lifetime, we see dramatic increases in stellar radius, by a factor greater than 3 for the 50\% case.
The evolution of our 2M$_\odot$ models is shown in Figure \ref{fig:evol2}. Feedback has no significant effect on the bulk evolution of these stars.

At the onset of the Dyson sphere feedback, a 1M$_\odot$ star is primarily radiative (out to 98\% of its mass) with a convective envelope. The 2M$_\odot$ star has an inner convective core, while the exterior is fully radiative. In the 1M$_\odot$ star, we see that the convective envelope carries some of the reflected energy into the stellar interior. The radiative core gets a slight cooling effect, slightly slowing fusion, while the envelope dramatically expands, radically changing the star's radius. With the 2M$_\odot$ star's lack of an outer convective region, very little of the reflected energy is able to reach the interior of the star, and the evolution of its bulk properties is essentially unaffected.

\section{Observational Characterization} \label{sec:obs}

To characterize the combined star-Dyson sphere system, we combine the spectra of the star and Dyson sphere into a system spectrum, from which we can calculate absolute magnitudes. We assume that the Dyson spheres emit waste heat as blackbodies at their effective temperatures. We compute this system as the sum of three components: the star, the interior of the Dyson Sphere, both of which are obscured by the Dyson sphere, and the exterior of the sphere, which is unobscured. We need to know the fraction of light that immediately escapes the system from these three sources, and the balance of radiation between the interior and exterior surfaces of the sphere. We begin with the full system flux equation 44 of \cite{dsreview}:
\begin{equation} \label{eq:review44}
    \Phi_\gamma = \frac{\pi}{d^2}(k_1 \Phi_{*,\gamma} + k_2 B_\gamma(T_\mathrm{int,eff}) + k_3 B_\gamma(T_\mathrm{ext,eff})),
\end{equation}
where $d$ is the distance to the system, $\Phi_{*,\gamma}$ is the star's specific intensity, $B_\gamma(T)$ is the Planck function for the Dyson sphere surface temperatures, and $k_i$ values are scaling factors we must calculate for each configuration. We calculate the luminosity of each component, scale them by the appropriate factors, and apply bolometric corrections to calculate absolute magnitudes. The bolometric corrections for the stars come from {\tt MESA} model output, and those for the Dyson spheres come from a modified version of {\tt blackbody.py} from {\tt mesa-r12778}'s {\tt colors} module \citep{mesa4}.

For each  of our stellar masses we apply our set of {\tt MESA} magnitudes for stars with no feedback and those with feedback levels between 1-50\%. In addition to these, we calculate the observable properties of Dyson spheres with lower $f$ values, between  $10^{-6} - 0.003$. These low $f$ values have no significant effect on the star itself, so we use our {\tt MESA} model with no feedback as the stellar components of these systems.

\subsection{The Stellar Component}
\citetalias{mesa1} implements bolometric corrections (BCs) to estimate absolute magnitudes in user-specified filter systems \citep{mesa4}. The BCs are calculated by linear interpolation over tables of $\log(T_{\rm eff})$
, $\log(g)$
, and [M/H]. 
We input BC tables from {\tt MESA} Isochrones and Stellar Tracks \citep[MIST,][]{mist1, mist2} for the {\it Gaia} 
and {\it WISE} filter systems. 

We select one point in the evolution of each star as our characteristic luminosity and bolometric correction values. For the 1 and 2M$_\odot$ stars, we select the point in time at which the star is halfway between the onset of its Dyson sphere and the end of its main sequence. For the 0.2 and 0.4M$_\odot$ stars, whose lifetimes are significantly longer than the age of the universe, we select the age of 10 Gyr, as their properties do not evolve significantly between the Dyson sphere's onset and the age of the universe.

\subsection{The Dyson Sphere Component}
So far, the only aspect of the Dyson spheres that we have defined is the fraction of luminosity reflected back onto the star ($f$). To characterize the structure in an observable way, we apply the AGENT formalism of \cite{ghat2} and \cite{dsreview}. The AGENT formalism is named for the five defining parameters of the Dyson sphere system (all powers normalized by the star's power output): the power of starlight intercepted $\alpha$, the power produced from other sources (e.g. fossil fuels) $\epsilon$, the power of the Dyson sphere's thermal waste heat $\gamma$, the power of other waste energy disposal $\nu$, and the characteristic temperature of the Dyson sphere's waste heat $T_{\rm waste}$. In this work, we set $\epsilon=\nu=0$, assuming that the structure does not generate its own energy or emit significant low-entropy emission (e.g. radio transmissions). 

We adopt a spherical shell model, centered over a star of radius $R_*$, with radius $R$ and Bond albedo $a$. Photons leaving the interior of the Dyson sphere have 3 possible fates: reflection, transmission, and absorption. We parametrize the fractional portions of radiation for each of these fates with:
\begin{equation}
    a+t+e=1,
\end{equation}
where $a$, $t$, and $e$ are the fractions of photons that get reflected, transmitted, and absorbed, respectively. For our feedback effect, we also need the parameter $s$, representing the probability that a photon emitted from or reflected by the interior of the sphere in a random direction will not immediately strike the star:
\begin{equation}
    s = \sqrt{1-(R_*/R)^2}.
\end{equation}
We also assume that the structures radiate heat equally on the interior and exterior (i.e.\ there is no strong thermal management), indicated as $\zeta = 1/2$ in the AGENT formalism. 

We examine 2 limiting cases for the $a$, and $e$ parameters. The first case, which we refer to as hot Dyson spheres, assumes black material, adopting a Bond albedo $a=0$. The second case, which we call cold Dyson spheres, corresponds to a spherical, specularly reflecting mirror that returns all light it collections back to the star. Such material absorbs no heat, so $e=0$.  Intermediate cases will have properties that are a compromise between these extremes. In both cases, we vary $t$, which corresponds to the transmittance of the sphere, which can be interpreted as the fraction of the star's solid angle the sphere does not cover. Since our models are 1D, this is not completely appropriate for stellar engines with large deviations from spherical symmetry.  

\subsubsection{Case 1: Hot Dyson Spheres}
For the hot Dyson sphere case, we assume that the sphere's material is black, so its Bond albedo is $a=0$. We assume that the sphere's transmissivity is equivalent at optical and infrared wavelengths, corresponding to opaque absorbers covering a fraction $e=1-t$ of the star.

Applying these assumptions to the AGENT formalism, we calculate the parameters that describe a system's observability, $\alpha$ and $\gamma$. We start by filling in our assumptions in the equations for the fractions of photons from the star and sphere that end up absorbed by the star, absorbed by the sphere, or escaped, in the limit of purely diffuse reflection from the shell, based on Table 2 from \cite{dsreview}. Our version is shown in Table \ref{tab:fs}.

\begin{table}[]
    \centering
    \caption{The fractions of stellar and Dyson sphere photons that end up being absorbed by the star, absorbed by the sphere, or escaping. This is the recreation of Table 2 of \citeauthor{dsreview} for the two configurations we explore: hot black spheres and cold mirrored spheres.}
    \begin{tabular}{c|c c|c c}
         & \multicolumn{2}{c|}{Starlight (*)} & \multicolumn{2}{c}{Thermal Emission from Sphere (s)} \\
         & hot & cold & hot & cold \\
         \hline
        Absorbed by Star (*) & 0 & 1-t & $\frac{1}{2} (1-s)$ & - \\
        Absorbed by Sphere (s) & 1-t & 0 & $\frac{1}{2} s (1-t)$ & - \\
        Escape (e) & t & t & $\frac{1}{2} (1 + s t)$ & -
    \end{tabular}
    \label{tab:fs}
\end{table}

We apply these to Equations 32-33 in \cite{dsreview} to calculate luminosity ratios required for energy balance:
\begin{equation}
    \frac{L_*}{\widetilde{L}} = \frac{2-s(1-t)}{1+t},
\end{equation}
\begin{equation} \label{eq:LsLtilde}
    \frac{L_{\rm s}}{\widetilde{L}} = \frac{2(1-t)}{1+t},
\end{equation}
\noindent where $L_*$ is the star's total luminosity, $\widetilde{L}$ is the star's luminosity due to power generated in its core (equivalent to L[2] used in {\tt MESA} and the above sections), and $L_{\rm s}$ is the luminosity of the Dyson sphere. We then plug these into Equations 34-35 to calculate our AGENT observability parameters:
\begin{equation} \label{eq:gamma_hot}
    \alpha = \gamma = \frac{(1-t)(1+st)}{1+t}
\end{equation}
where $\alpha$ is the fraction of the star's nuclear luminosity which does not escape the system as starlight and $\gamma$ is the fraction that ultimately escapes the system as thermal emission from the Dyson sphere. From our above equations, our feedback parameter relates to this formalism as:
\begin{equation}
    f = \frac{(1-s)(1-t)}{1+t}.
\end{equation}

For each of our stars at their selected point in time, we calculate a Dyson sphere's temperature and size for each of our feedback fractions $f$, for $t$ values from 0 to 0.99 (or the maximum possible for non-negative $s$ values.) From equations 37-40 in \cite{dsreview}, we calculate the temperature $T_{\rm s}$ of the Dyson sphere. Using our $L_{\rm s}$ from Equation \ref{eq:LsLtilde}, our effective temperature is:
\begin{equation}
    T_{\rm s} = \left( \frac{(1-s^2) \tilde{L}}{4\pi (1+t) \sigma R_*^2} \right)^{1/4},
\end{equation}
where $R_*$ is the star's radius. For this configuration, our Dyson sphere has the same internal and external temperature, so we can rewrite Equation \ref{eq:review44}, given Equations 45-47 of \cite{dsreview} as:
\begin{equation} 
    \Phi_\gamma = \frac{\pi}{d^2}(t R^2_* \Phi_{*,\gamma} + (1-t)(1+st) R^2_\mathrm{s} B_\gamma(T_\mathrm{s,eff})).
\end{equation}

\subsubsection{Case 2: Cold, Mirrored Dyson Spheres}
For case 2, we examine the opposite extreme, where the Dyson sphere is a spherical mirror returning starlight to the star without heating up. The Dyson sphere itself does not then have any significant temperature or luminosity. This corresponds to $e=0$ and $a=1-t$ in our equations. Photon fates in this case, assuming purely specular reflection are shown in Table \ref{tab:fs} with those of case 1.

For this case, our energy balance luminosity equations become:
\begin{equation}
        \frac{L}{\widetilde{L}} = \frac{1}{t},
\end{equation}
\begin{equation} \label{eq:LsLtilde2}
    \frac{L_{\rm s}}{\widetilde{L}}  = 0.
\end{equation}
Our AGENT observability parameters are:
\begin{equation} \label{eq:alpha_cold}
    \alpha = \gamma = 0 .
\end{equation}

With no Dyson sphere luminosity, the luminosity poured onto our star is simply the fraction of its light reflected:
\begin{equation}
    f = a = 1-t .
\end{equation}
For each of our stellar models, we just see the luminosity of the star scaled by the factor $t$, with no additional thermal component, so our total luminosity equation is simply:
\begin{equation}
    \Phi_\gamma = \frac{\pi}{d^2} t R_*^2 \Phi_{*,\gamma}.
\end{equation}

\section{Discussion}
\subsection{Color-magnitude Diagrams}

\begin{figure}
    \centering
    \includegraphics[width=0.9\textwidth]{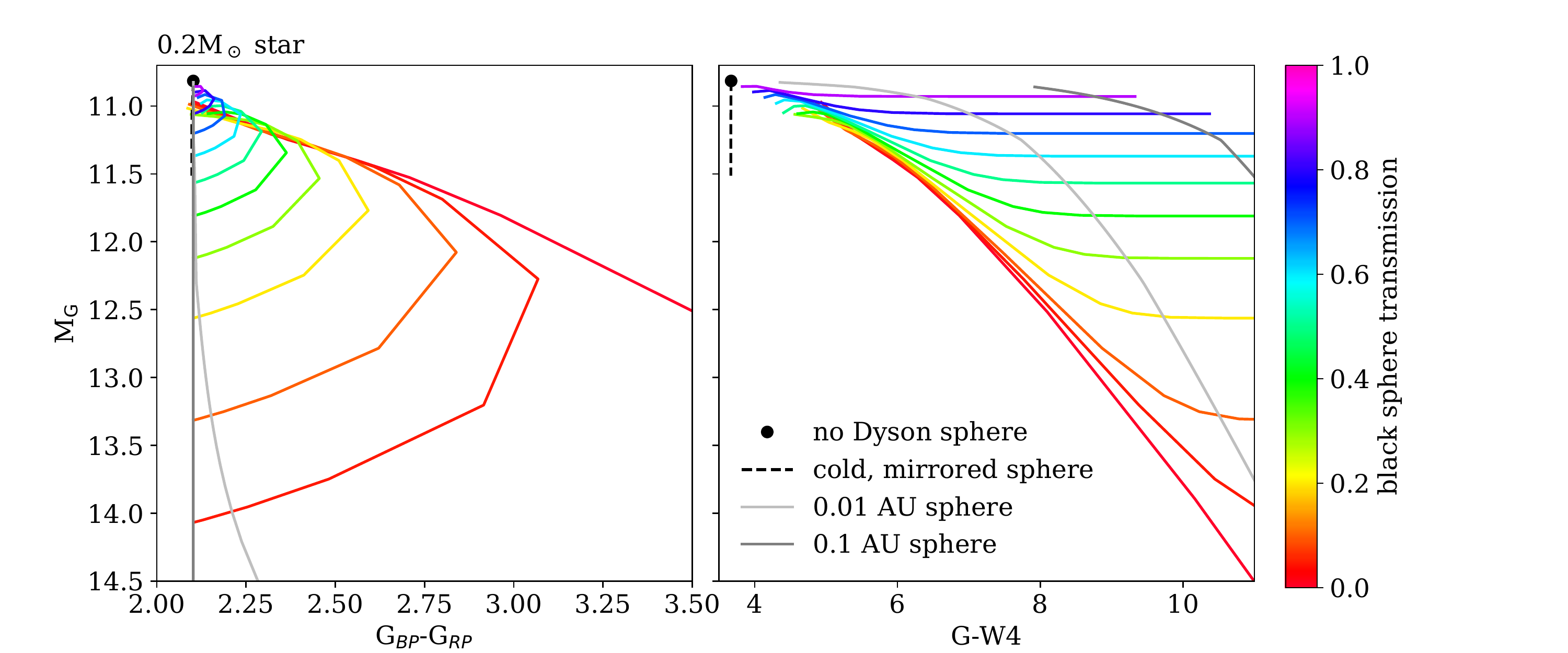}
    \includegraphics[width=0.9\textwidth]{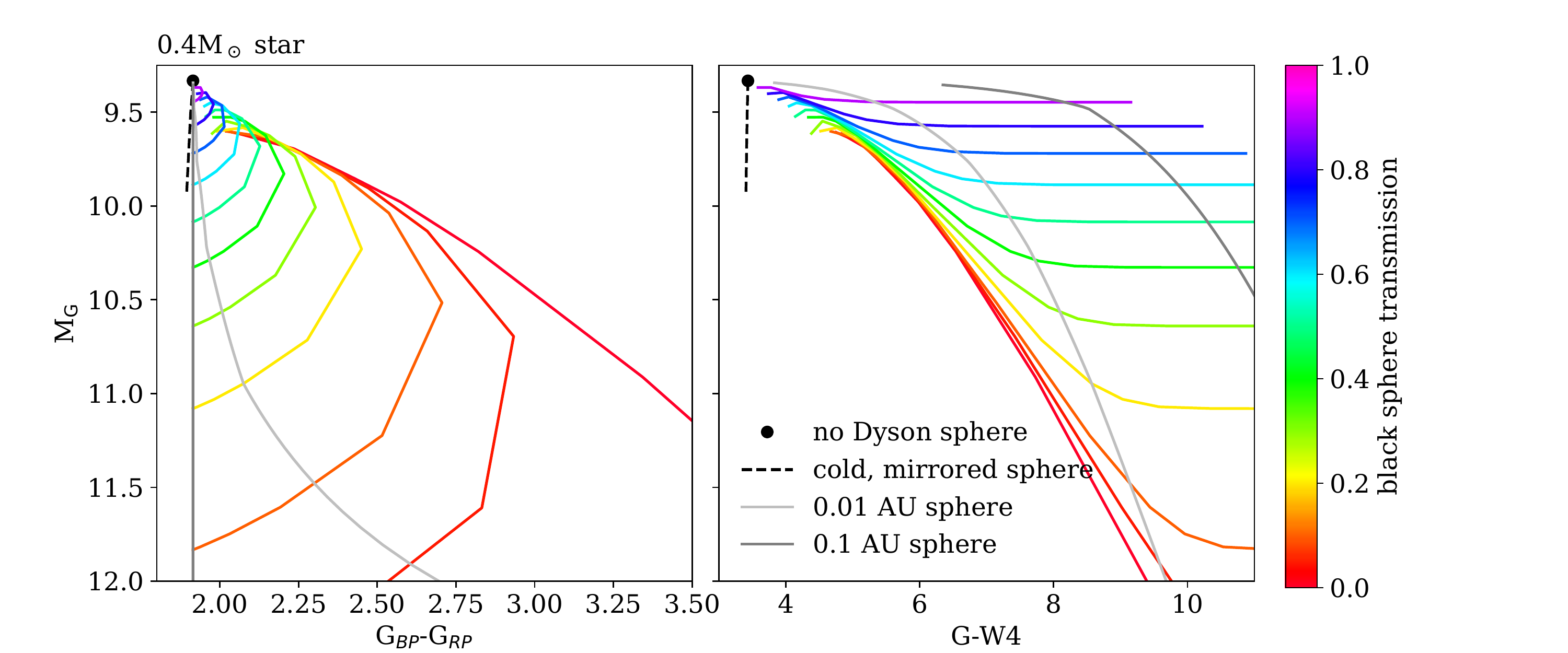}
    \caption{Low-mass star color-magnitude diagrams for combined star and Dyson sphere systems. Top: 0.2M$_\odot$, Bottom: 0.4M$_\odot$. The bare star is shown as a black dot, and the extending black dotted line shows a range of cold, reflective Dyson spheres from $f=0$ to $f=0.50$. Colored lines are drawn at intervals from $t=0$ to $t=0.99$ to show the hot Dyson sphere models. Light gray marks a Dyson sphere at radius 0.01 AU and medium gray at 0.1 AU. }
    \label{fig:lm_cmds}
\end{figure}

\begin{figure}
    \centering
    \includegraphics[width=0.9\textwidth]{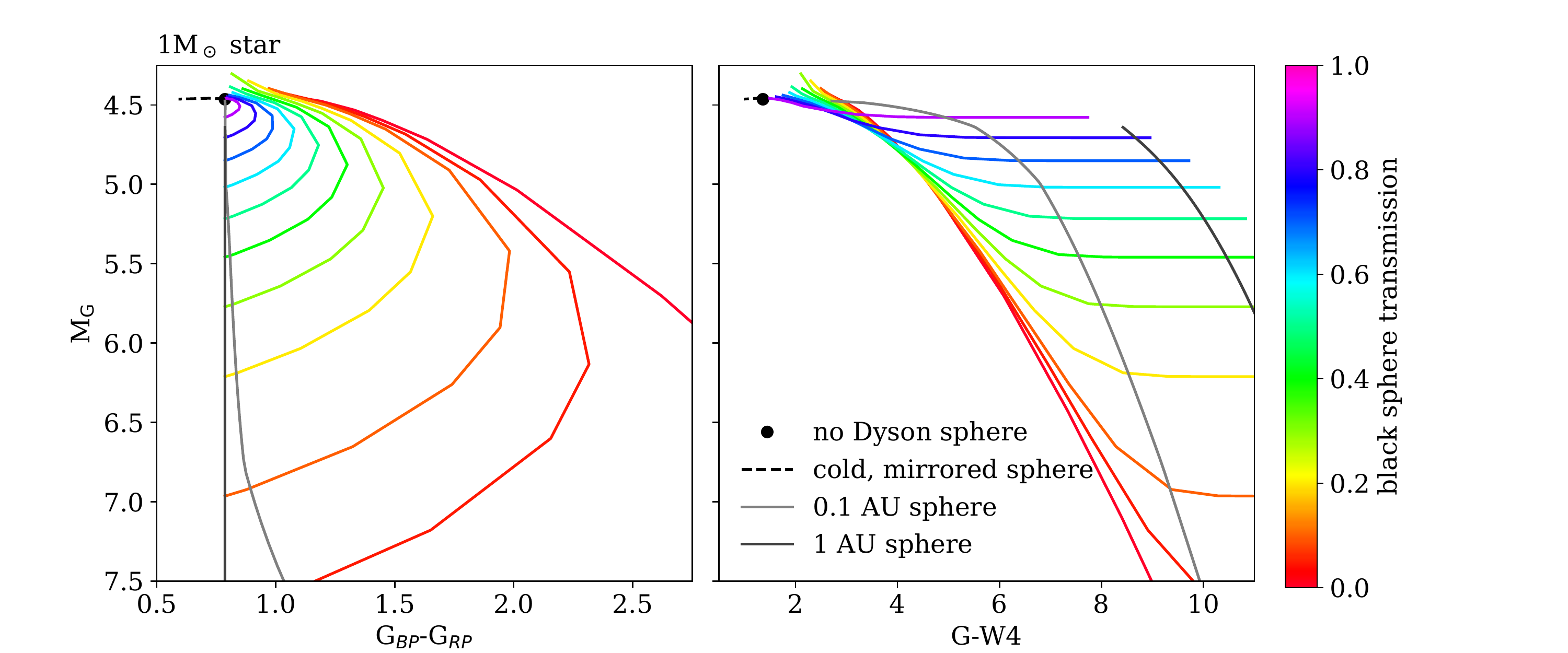}
    \includegraphics[width=0.9\textwidth]{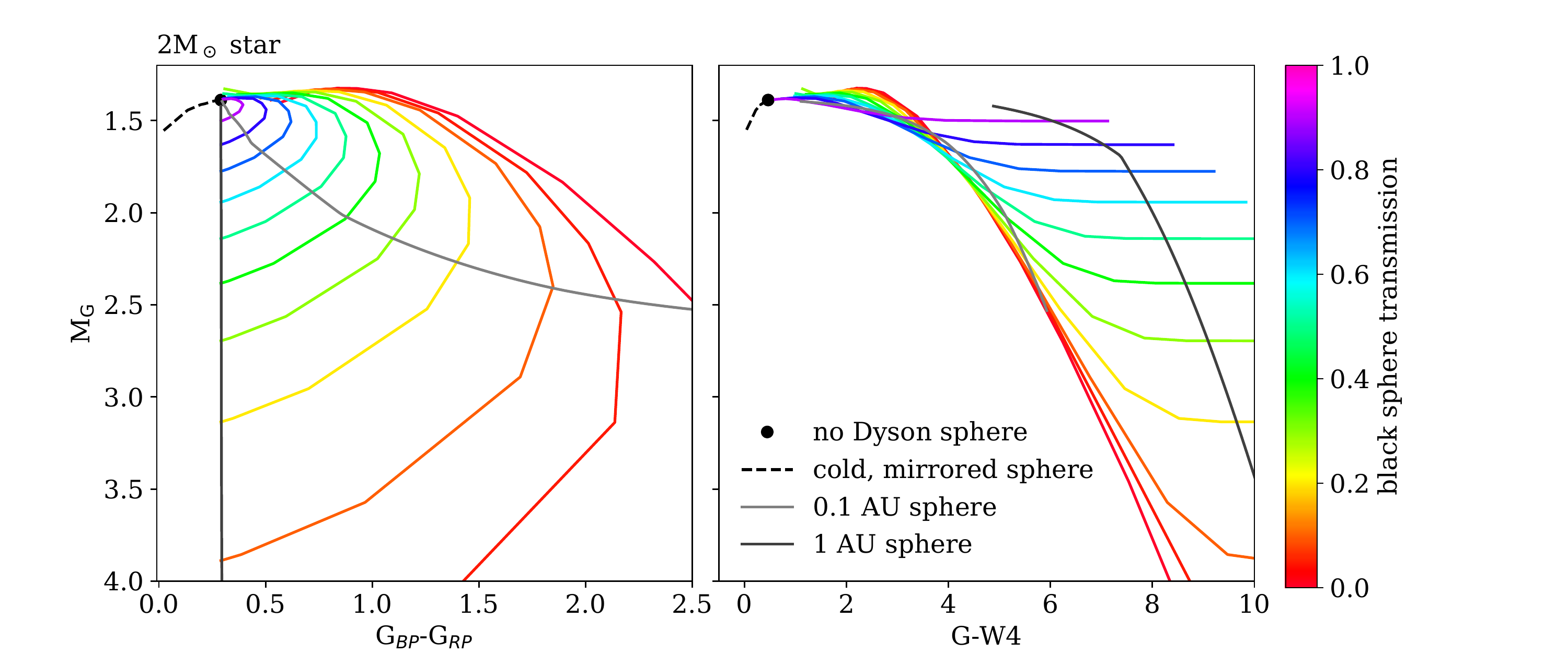}
    \caption{Intermediate-mass star color-magnitude diagrams for combined star and Dyson sphere systems. Top: 2M$_\odot$, Bottom: 2M$_\odot$. The bare star is shown as a black dot, and the extending black dotted line shows a range of cold, reflective Dyson spheres, from $f=0$ to $f=0.50$. Colored lines are drawn at intervals from $t=0$ to $t=0.99$ to show the hot Dyson sphere models. Medium gray marks a Dyson sphere at radius 0.1 AU and dark gray at 1 AU. }
    \label{fig:hm_cmds}
\end{figure}

We take a closer look at selected {\it Gaia} and {\it WISE} color-magnitude diagrams (CMDs) shown for low-mass stars in in Figures \ref{fig:lm_cmds} and intermediate-mass stars in \ref{fig:hm_cmds}. From {\it Gaia}, we use the G (green), G$_{\rm BP}$ (blue), and G$_{\rm RP}$ (red) filters to characterize the systems at optical wavelengths. From {\it WISE}, we use the W4 (22 $\mu$m) filter to include an infrared component.

In each CMD, we show the bare star and a line tracing cold, mirrored sphere systems with reflection fractions from 0 to 0.50. We also show a series of curves tracing the magnitudes of hot, non-reflective Dyson spheres with different transmission levels. Along these curves, the feedback parameter $f$ can range from 10$^{-6}$ to 0.50, though unphysical $f$ and $t$ combinations which would require a Dyson sphere smaller than its host star are excluded. In addition, we trace select lines of constant hot Dyson sphere radius. The data behind the figures are available in machine-readable form.

For the hot-to-warm Dyson sphere configurations, high values of $f$ begin in the upper left of each CMD and extend downward toward lower $f$. For each stellar mass, we see similar general trends. In the left-side diagrams, moving toward higher $f$, we see the stars dim in absolute G magnitude, as they redden then bluen again in G$_{\rm BP}$-G$_{\rm RP}$. This happens because at very high feedback levels, the Dyson sphere's temperature is very close to that of the star. As feedback decreases, we see the Dyson sphere begin to cool, reddening the system. At very low feedback levels, the Dyson spheres become very cool and no longer contribute significantly to optical magnitudes, so we begin to see the star's natural color again. On the right-side figures, we see that each constant $t$ line generally dims and reddens as we move toward lower feedback levels. We see continuous reddening here as we move to cooler and cooler Dyson spheres, as the G-W4 color can pick up the infrared output of these relatively low-temperature objects.

Moving from high $t$ lines to lower, we see this effect exaggerated, as the Dyson spheres intercept more starlight and contribute more to the overall appearance. As the classical idea of a Dyson sphere, we can examine a solar mass star with low transmission of starlight through the sphere and a Dyson sphere radius of roughly 1 AU. We see that the feedback levels are very low and that the systems will appear, relative to a bare solar mass star, to be dimmed in the optical range and reddened in both optical and infrared colors.

Next, we examine the CMDs for cold, mirrored spheres. Low feedback levels (high transmission values) begin at the bare star point. For the 0.2M$_\odot$ star, we see dimming in the G band and no significant change in color, as the star's effective temperature does not significantly change, and there is no Dyson sphere emission. For the 0.4M$_\odot$ star, we see dimming in the G band and the star getting slightly bluer, as the star very slightly increases in effective temperature. The 1M$_\odot$ star does not significantly dim in the G band and becomes bluer. Feedback significantly increases its effective temperature, making it bluer, but not appearing brighter due to decreased transmission. The 2M$_\odot$ star dims and becomes bluer. Its spectrum peaks in UV wavelengths, so the effective temperature increase is not enough to counteract the decreased transmission by the higher coverage mirror systems.

\subsection{When Feedback Matters}
The feedback of energy back a stellar surface resulting from a warm and/or reflective Dyson sphere can strongly affect the appearance of a star under certain circumstances, particularly low-mass stars and high feedback $f$ values. For low mass stars, we have demonstrated that nuclear luminosity is significantly affected. To cause a 1\% change in the star's nuclear luminosity, 0.2 and 0.4 M$_\odot$ stars require 1.2\% and 1.3\% feedback, respectively. Interestingly, we see a partial cancellation in the effects on the star's effective temperature. Feedback increases the temperature of the star's exterior, increasing luminosity, while the reduction in nuclear burning decreases luminosity. Ultimately, their effective temperatures do not change noticeably for the feedback levels explored ($\leq$50\%).

For high mass stars, feedback cannot penetrate far into the star, so a very large amount is required to affect nuclear burning, often above the limit of 50\% explored here. To produce a 1\% change in the 1M$_\odot$ star's nuclear luminosity, 45\% feedback is required. None of our models were able to significantly change the 2M$_\odot$ star's nuclear burning. Since there is not a significant cooling effect in these stars, we see that their effective temperatures can be significantly impacted. For a 1\% change in effective temperature for 1 and 2 M$_\odot$ stars, feedback levels of 5.8\% and 5.5\%, respectively, are required. For a cold, mirrored sphere, for example, these would be the fraction of the star's solid angle ($f = a = 1-t$) required to be covered in mirrors for the feedback to significantly affect stars' effective temperatures.

\begin{figure}
    \centering
    \includegraphics{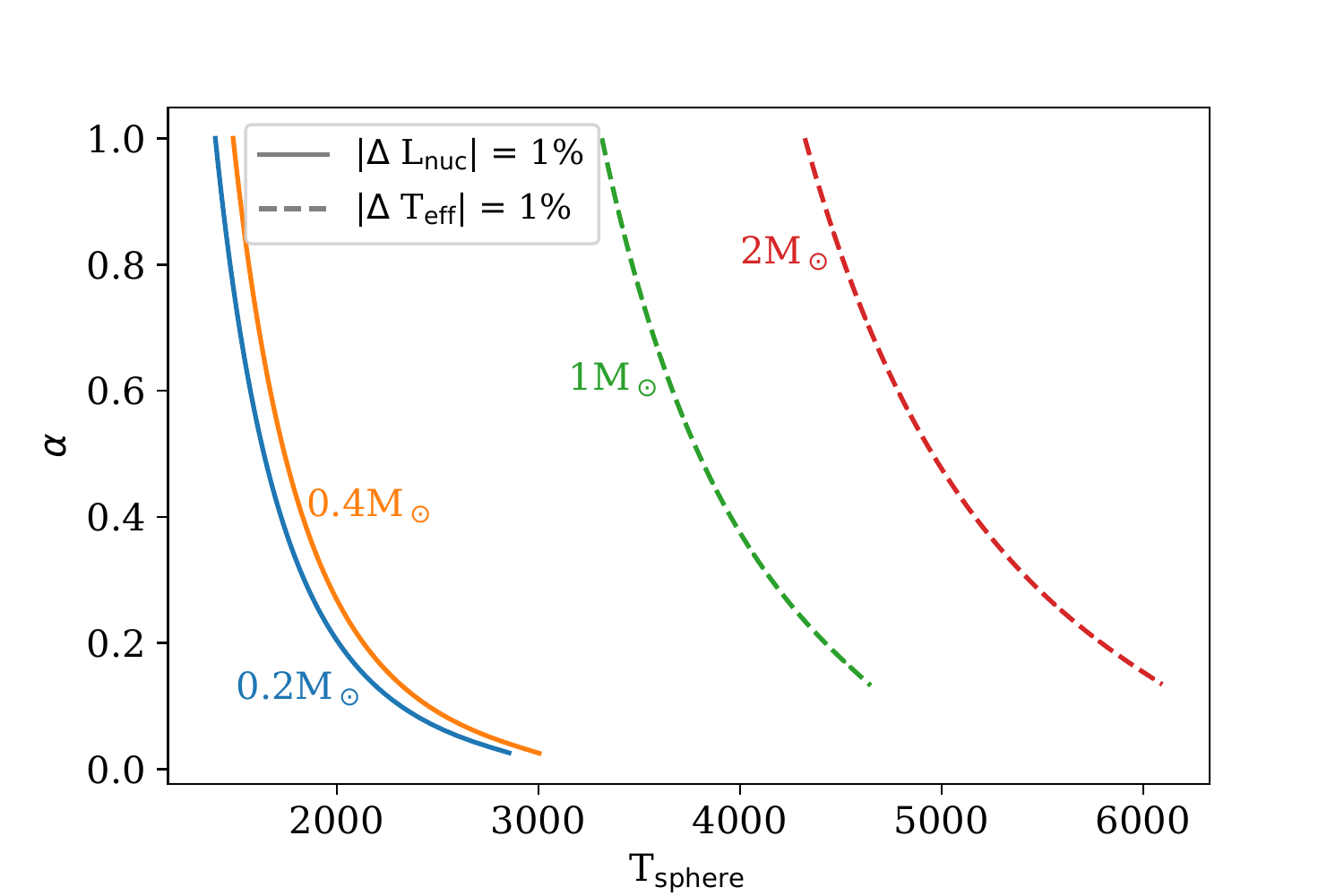}
    \caption{Lines of T$_{\rm sphere}$ versus $\alpha$ for hot Dyson spheres at which 1\% changes in stellar properties occur. For the low mass stars, we trace the range of possible sphere properties at which L$_{\rm nuc}$ changes by 1\%, as stellar T$_{\rm eff}$ does not significantly change. For the higher mass stars, L$_{\rm nuc}$ does not significantly change, so we trace the limit for T$_{\rm eff}$ to change by 1\%. The changes become stronger as one moves toward the upper right of the lines (hotter spheres that convert more starlight). Ultimately, none of the stars change significantly in nuclear burning or effective temperature for the typically assumed Dyson sphere temperatures of a few hundred Kelvin. }
    \label{fig:alpha_ts}
\end{figure}

To demonstrate how these feedback levels translate to physical properties of hot Dyson spheres, we plot $\alpha$ (the fraction of stellar luminosity which does not escape as starlight) and $T_{\rm sphere}$ for each stellar mass at the $f$ values that cause significant changes to stars' appearances in Figure \ref{fig:alpha_ts}. Ultimately, these significant observable changes only occur for Dyson spheres on the order of a few thousand Kelvin and not for those of typically assumed Dyson sphere temperatures of a few hundred Kelvin.

\subsection{Stellar Engineering}

We have limited our analysis to $f<0.5$ because this represents an extreme outer limit to what might be expected from Dyson Spheres used as energy collectors, and in part because higher values required additional modifications to {\tt MESA} to capture the physics correctly. 
But higher values, even up to $f=1$, might be interesting to consider as components of a stellar engineering project. As we have shown, returning starlight to a star can increase its lifetime, especially if a significant fraction of the star's outer mass is convective. 

Whether such a project could succeed on a star with a substantial radiative component is unclear, but in principle ``bottling up'' a star completely should unbind it on a Kelvin-Helmholtz timescale and quench the nuclear activity in its core. Such a project might be desirable and thus be done intentionally, either to extend the star's life, prevent it from experiencing post-main-sequence evolution, or even extract its mass. Exploring such possibilities is a topic for future work.

\section{Conclusion} \label{sec:concl}
Irradiated stars expand and cool. A Dyson sphere may send a fraction of a star's light back toward it, either by direct reflection or thermal re-emission. This returning energy can be effectively transported through convective zones but not radiative zones. So, it can have strong impacts on low mass main sequence stars with deep convective zones which extend to the surface. It causes them to expand and cool, slowing fusion and increasing main sequence lifetimes. For higher mass stars with little to no convective exterior, the returned energy cannot penetrate far into the star and therefore has little effect on the star's structure and evolution, besides some surface heating.

We have used \citetalias{mesa1} to model the structure and evolution of stars with masses from 0.2 to 2 M$_\odot$ with returned luminosity fractions from 0.01 to 0.50. We have incorporated the effects of feedback on stars and calculated Dyson sphere properties using the AGENT formalism of \cite{ghat2} and the feedback equations of \cite{dsreview}. We have compiled absolute magnitudes in {\it WISE} and {\it Gaia} filters for a variety of combined star-Dyson sphere systems. These are shown in color-magnitude diagrams in this work and are also available as machine-readable tables. 

For our 0.2 and 0.4 M$_\odot$ stars, feedback levels above roughly 1\% cause at least a 1\% change in nuclear luminosity; their effective temperatures do not significantly change. For our 1 and 2 M$_\odot$ stars, feedback levels above roughly 6\% cause at least a 1\% change in the star's effective temperature; their nuclear luminosities do not significantly change. Physically, these limits may correspond with a cold, mirrored surface covering the specified fraction of the star's solid angle. For light-absorbing, non-reflective Dyson spheres, these feedback levels correspond to very hot spheres, with temperatures of thousands of Kelvin. This is well above the few hundred Kelvin temperatures typically assumed in Dyson sphere studies, but such hot spheres have been considered in the literature before \citep{Osmanov2018}.

Although the circumstances under which this feedback affects stellar structures are rather extreme, we have demonstrated that it can have a significant effect on the observable properties of certain systems, including mirrored stellar engines and hot Dyson spheres. Mirrored spheres might be relevant for stellar engineering projects designed to extend a star's lifetime, reduce its luminosity, or extract its mass.

We have also compiled expected absolute magnitudes for Dyson spheres at a wide range of feedback levels, including at significantly lower, perhaps more likely levels of feedback, to help guide future searches.

\begin{acknowledgements}
We thank Noah Tuchow and Josiah Schwab for helpful advice regarding the {\tt MESA} modelling process. 

The Penn State Extraterrestrial Intelligence Center and Center for Exoplanets and Habitable Worlds are supported by the Pennsylvania State University, the Eberly College of Science, and the Pennsylvania Space Grant Consortium. This research has made use of NASA’s Astrophysics Data System Bibliographic Services.
\end{acknowledgements}

\vspace{5mm}

\software{ MESA \citep{mesa1,mesa2,mesa3,mesa4,mesa5}, Astropy \citep{astropy1, astropy2}, Matplotlib \citep{matplotlib}, NumPy \citep{numpy}, SciPy \citep{scipy}}
\typeout{} 
\bibliography{refs}
\bibliographystyle{aasjournal}

\end{document}